\newcommand{\op}{\mathcal{O}}

\newcommand{\me}[1]{\ensuremath{\langle \op_{#1}^q  \rangle}}

\documentclass{PoS}

\title{Lattice QCD Impact on Determination of CKM Matrix: Status and Prospects}

\ShortTitle{LQCD Impact on Determination of CKM Matrix}

\author{\speaker{Steven Gottlieb}\\
        Indiana University, Bloomington, IN 47405, USA\\
        E-mail: \email{sg@iu.edu}}

\abstract{
Lattice QCD is an important tool for theoretical input for flavor physics.
There have been four reviews by the Flavour Lattice Averaging Group (FLAG).
This talk will review the current status of the magnitude of eight of the 
nine CKM matrix elements, borrowing heavily from the most recent FLAG review
(co-authored by the speaker).  Future prospects for improving the determination
of the CKM matrix will be discussed.}

\FullConference{37th International Symposium on Lattice Field Theory - Lattice2019\\
		16-22 June 2019\\
		Wuhan, China}

\begin{document}

\section{Introduction}
I was asked to ``review very recent FLAG results on standard model parameters
and renormalization.''  This is a very broad charge and one that is ill-suited
to the amount of time available, so I will restrict my attention to results
from lattice QCD calculations that have an impact on the determination of
the CKM mixing matrix.  Even with this restriction, it will be necessary to
summarize results at a rather high level.  Many details of the calculations
can be found in the latest and recent FLAG 
reviews~\cite{Aoki:2016frl,Aoki:2019cca}, and, of course, the 
original papers cited therein.

The Cabibbo-Kobayashi-Maskawa (CKM) mixing matrix is fundamental to the
field of flavor physics within the Standard Model (SM) of Elementary Particle and
Nuclear Physics.  Kobayashi and Maskawa were awarded the Nobel Prize for
their realization that with three generations of quarks the matrix may contain a
complex phase that results in $CP$ violation.  Although it seems that this
is not sufficient to explain the baryon asymmetry of the universe, there
are many processes in which to test the CKM scenario, and any such tests
in which the SM fails to explain the observations could give a window into
new physics beyond the standard model (BSM).

A few words about my background may be in order.  I am a founding member of the 
MILC Collaboration and a member of the Fermilab Lattice/MILC effort that is 
some 16 years old.  I am also a member of the Flavour Lattice Averaging Group
(FLAG) where I have been working in the $B$ and $D$ semileptonic working
group.  However, this is not a FLAG approved talk, so I am solely responsible
for its content.  The two most recent editions of the FLAG report are
in Refs.~\cite{Aoki:2016frl} and \cite{Aoki:2019cca}.
I will use many plots from the most recent FLAG report
and cover results from several of the working groups.  I will also include 
several graphs from the Fermilab Lattice/MILC Collaborations.  I am 
grateful for the work of all my FLAG, Fermilab Lattice, and MILC collaborators.

\section{CKM Matrix}
The CKM matrix describes how quarks mix under the weak interaction, that is,
the misalignment of mass eigenstates and weak eigenstates.
The CKM matrix is shown in  expression~\ref{eq:ckmmatrix}.
The matrix elements are shown in bold type,
and beneath the elements in the first two rows you will find one or two weak
decays that can be used to determine that matrix element, if we can accurately
calculate the QCD contribution to the decay.  Under the last row of elements
are two hadronic matrix elements that remind us that $B_{(s)}$ mixing 
allows us to determine $V_{td}$ and $V_{ts}$.

        \begin{eqnarray}
        \left(
        \begin{array}{ccc}
        { \bf{V_{ud}}}   &  \bf{ V_{us}}  &   \bf{ V_{ub} }\\
        \textcolor{red}{\pi\to l\nu} &  \textcolor{red}{K\to l\nu} & \textcolor{red}{B\to l\nu}\\
                    & \textcolor{blue}{K\to\pi l\nu} & \textcolor{blue}{B\to\pi l\nu}\\
        \bf{ V_{cd} }  &  \bf{ V_{cs}  } &   \bf{ V_{cb}} \\
        \textcolor{blue}{D\to \pi l\nu} & \textcolor{blue}{D\to K l\nu} & \textcolor{blue}{B\!\to\! D^{(\!*\!)}\! l \nu} \\
         \textcolor{red}{D\to l\nu} &  \textcolor{red}{D_s\to l\nu} & \textcolor{blue}{\Lambda_b\to\Lambda_c l\nu}  \\
        \bf{ V_{td}}  &\bf{ V_{ts}}  & \bf{ V_{tb}} \\
        \textcolor{green}{\langle B_d | \overline{B}_d\rangle }&
        \textcolor{green}{\langle B_s | \overline{B}_s\rangle }\\
        \end{array}
        \right)
\label{eq:ckmmatrix}
        \end{eqnarray}

The CKM matrix is unitary so each row and each matrix is a complex unit
vector.  Each row (column) is orthogonal to the other two rows (columns).
Violations of unitarity are evidence of BSM physics.  It is
important to use multiple processes to determine each matrix element.  If
two different processes infer different values for the same CKM matrix
element, that would also be evidence for non-standard model physics.
Of course, we would very much like to have solid evidence for BSM physics,
but that requires precise determination of the standard model contribution.
Lattice QCD is one of the best tools for calculating those contributions.

As a first example, let's consider the leptonic branching fraction for 
the $D_{(s)}$ meson:
\begin{equation}
{\mathcal{B}}(D_{(s)} \to \ell\nu_\ell)= {{G_F^2|V_{cq}|^2 \tau_{D_{(s)}}}\over{8 \pi}} f_{D_{(s)}}^2 m_\ell^2 
m_{D_{(s)}} \left(1-{{m_\ell^2}\over{m_{D_{(s)}}^2}}\right)^2 \,.
\end{equation}
On the right hand side, the Fermi constant, lepton mass, meson mass, and meson
lifetime appear.  These are all well determined.  Also appearing are
the CKM matrix element $V_{cq}$ that we would like to determine, and the
so-called hadronic decay constant $f_{D_{(s)}}$ that we calculate using
lattice QCD.  A second example is a semileptonic $D$ meson decay for which the 
differential decay rate may be written:
\begin{eqnarray}
        \frac{d\Gamma(D\to P\ell\nu)}{dq^2} = \frac{G_{\rm\scriptscriptstyle F}^2 |V_{cx}|^2}{24 \pi^3}
        \,\frac{(q^2-m_\ell^2)^2\sqrt{E_P^2-m_P^2}}{q^4m_{D}^2} \,
        \bigg[ \left(1+\frac{m_\ell^2}{2q^2}\right)m_{D}^2(E_P^2-m_P^2)|f_+(q^2)|^2 & \nonumber\\
+ \frac{3m_\ell^2}{8q^2}(m_{D}^2-m_P^2)^2|f_0(q^2)|^2 & \!\!\!\! \bigg]\,, \label{eq:DtoPiKFull}
\end{eqnarray}
where $x = d, s$ is the daughter light quark, $P= \pi, K$ is the
daughter light-pseudoscalar meson, $q = (p_D - p_P)$ is the
momentum of the outgoing lepton pair, and $E_P$ is the light-pseudoscalar 
meson energy in the rest frame of the decaying $D$.  A similar formula holds
for other heavy-light mesons such a $D_s$, $B$ or $B_s$.  The hadronic
physics that we require is expressed in terms of the two form factors
$f_+(q^2)$ and $f_0(q^2)$.  They are defined in terms of the hadronic matrix
element of the flavor-changing vector current
$V_\mu = \overline{x} \gamma_\mu c$,
\begin{equation}
\langle P| V_\mu | D \rangle  = f_+(q^2) \left( {p_D}_\mu+ {p_P}_\mu - \frac{m_D^2 - m_P^2}{q^2}\,q_\mu \right) + f_0(q^2) \frac{m_D^2 - m_P^2}{q^2}\,q_\mu \,.
\end{equation}
The experimental observable depends upon known quantities,
a CKM matrix element, and the hadronic information that is encoded 
the form factors which depend upon $q^2$.

\begin{figure}[tb]
\begin{center}
\includegraphics[width=0.75\textwidth]{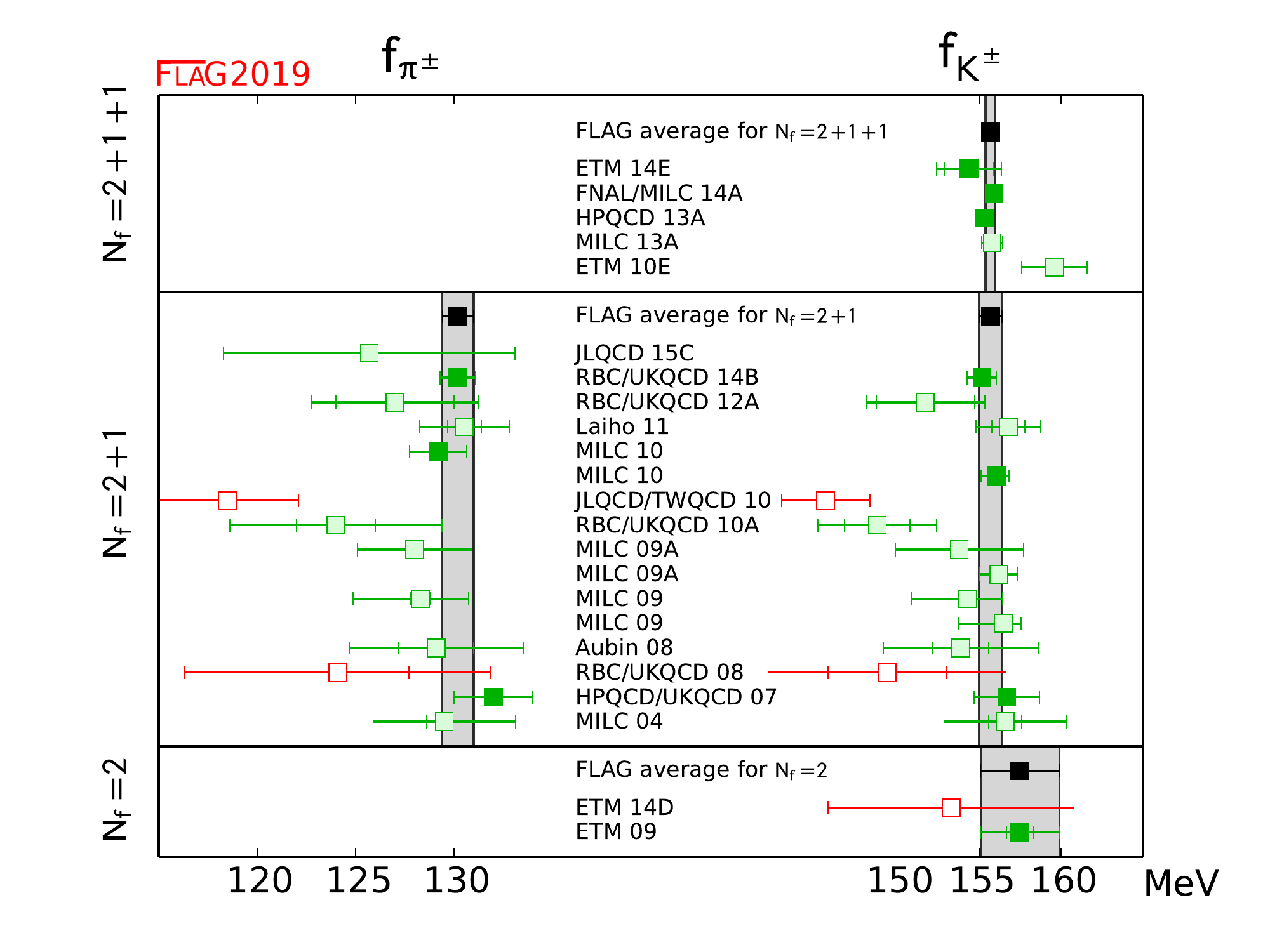}
\vspace{-5mm}
\caption{
FLAG 2019 compilation of results for $f_\pi$ and $f_K$.  Meaning of the
colors is explained the the text.}
\label{fig:FLAG_fKandfpi}
\end{center}
\end{figure}

\section{First Row}
Because $|V_{ub}|$ is so small,
we can test unitarity of the first row of the CKM matrix quite well (with
current precision) looking only at pion and kaon leptonic decays and kaon
semileptonic decay.  
Figure~\ref{fig:FLAG_fKandfpi} summarizes results for $f_\pi$ and $f_K$, the
pion and kaon decay constants, respectively.  This is the first of a number of
plots from FLAG~\cite{Aoki:2019cca},
so we should explain the color coding.  Solid green symbols
correspond to calculations for which there should be sufficient control
of systematic errors.  There are criteria for lattice volume, quark masses (to
control the chiral limit) and number of lattice spacings (to control the
continuum limit).  Some quantities have additional quality criteria detailed
in the FLAG report~\cite{Aoki:2019cca}.
Points in green with open plotting symbols
have been superseded, usually because a group has added additional ensembles
or statistics.  Points plotted in red are deemed not to have adequate
control of systematic errors.  The black points and gray bands
are the FLAG average values.  Only solid green points are included in the
FLAG average values.  Each graph is generally divided into sections depending
upon the number of dynamical quarks in the sea.  The values $N_f=2$, 2+1,
and 2+1+1 are used for two-light flavors, two light plus strange, and
two light plus strange plus charm, respectively.  In most cases, there is
a FLAG average for each value of $N_f$.  We will restrict our attention
to $N_f=2+1$ and 2+1+1.

\begin{figure}[tb]
\begin{center}
\includegraphics[width=0.7\textwidth]{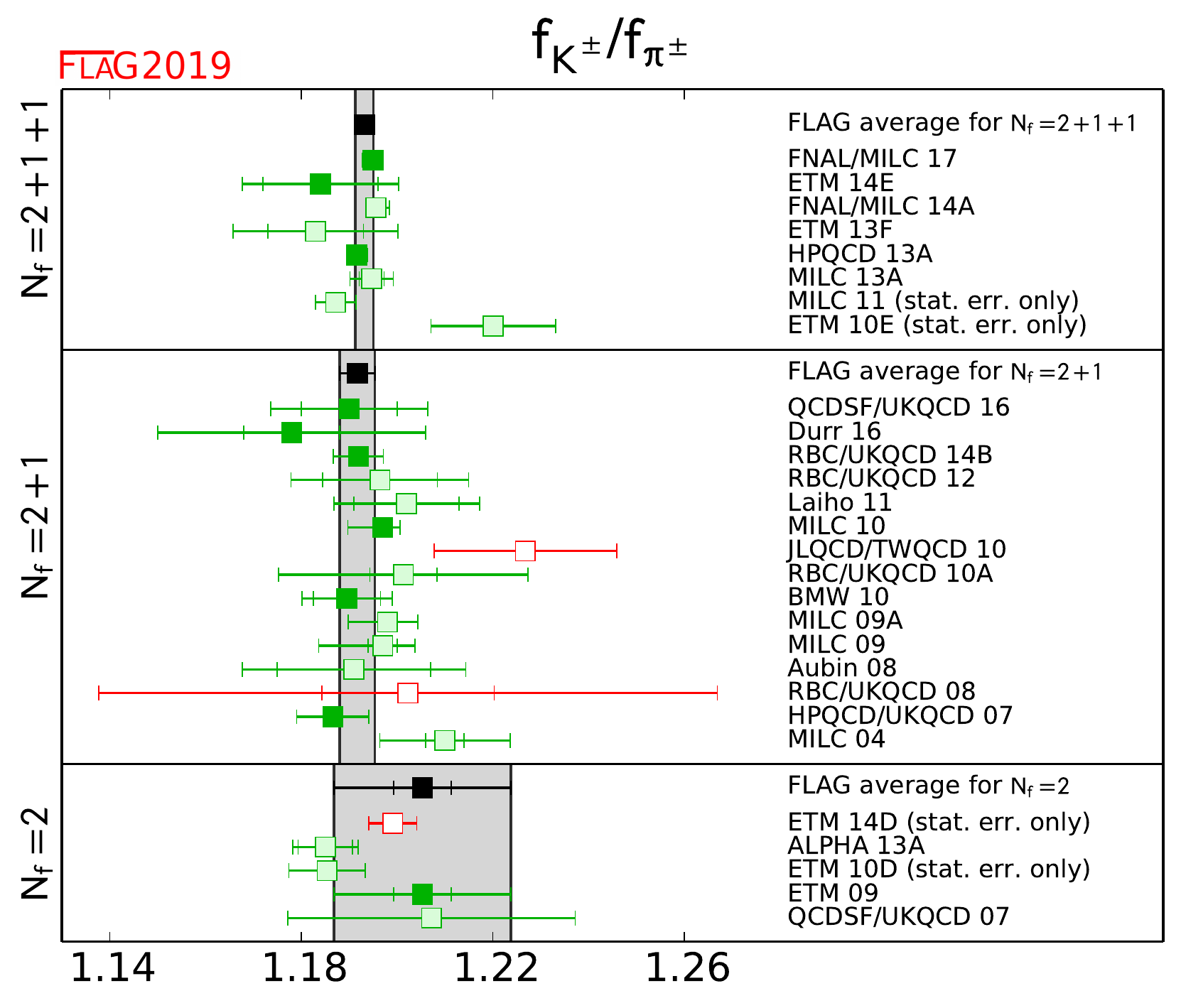}
\vspace{-5mm}
\caption{
\label{fig:FLAG_fKoverfpi}
FLAG 2019 compilation of ratio $f_K/f_\pi$.
Comparison of calculations of decay constant ratio $f_K/f_\pi$ with
$N_f=2+1+1$, $2+1$, and $2$ sea quark flavors.  From Ref.~\cite{Aoki:2019cca}.
}
\end{center}
\end{figure}
\begin{figure}[h!bt]
\begin{center}
\includegraphics[width=0.7\textwidth]{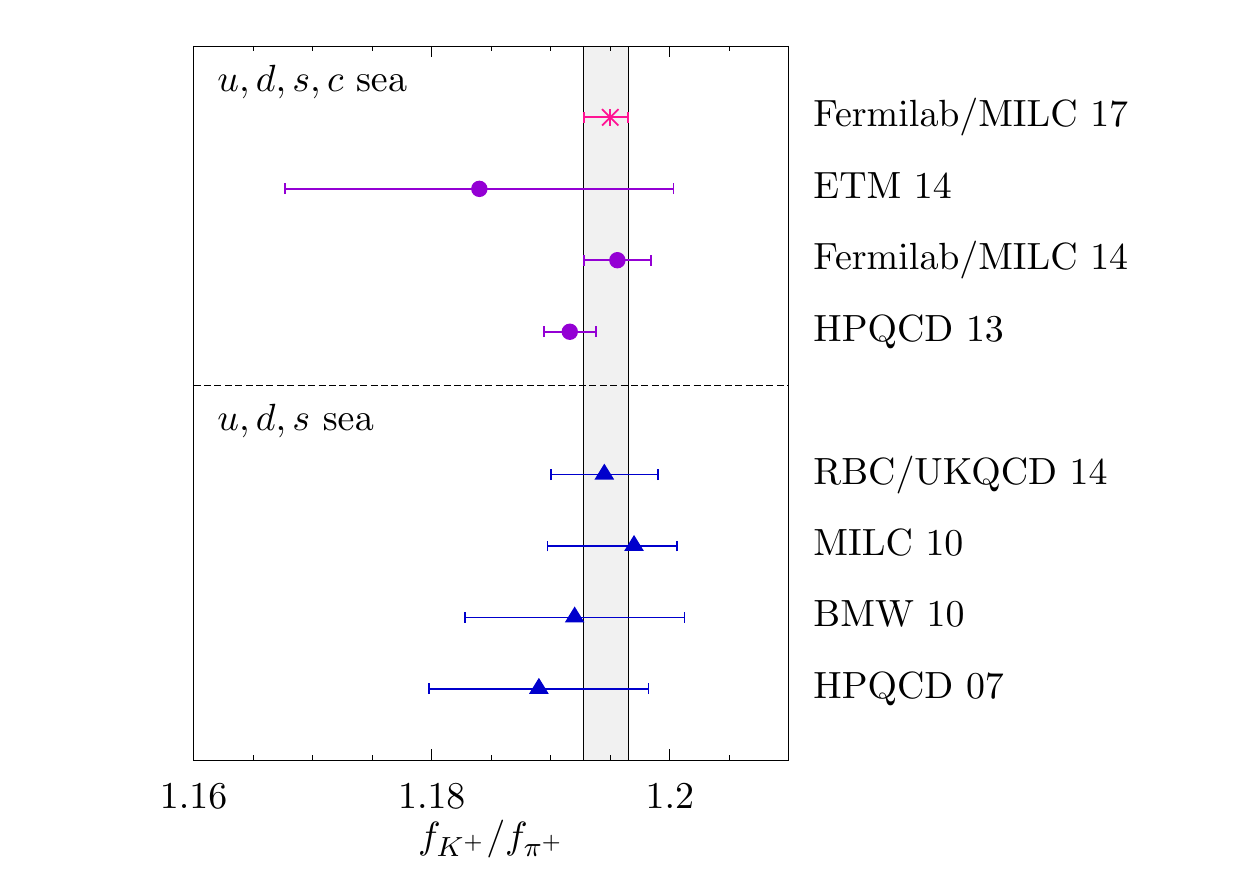}
\vspace{-5mm}
\caption{
\label{fig:FNALMILC_fKoverfpi}
Comparison of calculations of decay constant ratio $f_K/f_\pi$ with
$N_f=2+1+1$ and $2+1$ sea quark flavors.  From Ref.~\cite{Bazavov:2017lyh}.
}
\end{center}
\end{figure}

In Fig.~\ref{fig:FLAG_fKandfpi}, one notices that in many cases there is
a value of $f_K$, but none for $f_\pi$.  This is usually because $f_\pi$
has been used to set the scale.  Marciano~\cite{Marciano:2004uf}
pointed out that the ratio 
$f_K/f_\pi$ can be used to determine
the ratio $|V_{us}/V_{ud}|$ as leptonic decays of both pion and kaon
are well measured in experiment.  
As summarized by the Particle Data Group~\cite{Tanabashi:2018oca}
and Moulson at CKM 2017~\cite{Moulson:2017ive},
\begin{equation}
\left|{V_{us}\over V_{ud}}\right| {f_{K^\pm}\over f_{\pi^\pm}} = 0.2760(4) .
\end{equation}
Thus, the experimental error is 0.15\%.  Figures~\ref{fig:FLAG_fKoverfpi}
and \ref{fig:FNALMILC_fKoverfpi} display summaries of results from
multiple groups for the
decay constant ratio. Figure~\ref{fig:FLAG_fKoverfpi} is from 
FLAG~\cite{Aoki:2019cca}, while Fig.~\ref{fig:FNALMILC_fKoverfpi}
is from a recent Fermilab/MILC publication~\cite{Bazavov:2017lyh}.
The former is more comprehensive,
but the latter uses a finer scale, and one can actually see the error bars
on the most precise calculations, as they are no longer obscured by the
plotting symbols.  The FLAG 2019 result for $N_f=2+1+1$ is 1.1932(19).
The previous FLAG result~\cite{Aoki:2016frl} was 1.193(3).  
The FNAL/MILC 2017 result is 
$1.1950(^{+15}_{-22})$.  The theory error has been reduced to 0.16\%,
quite comparable to the experimental error.

Let us now turn from leptonic decays to the semileptonic decay of the
K meson.  Since semileptonic decays have three-body final states,
the kinematics are slightly more complicated, and there is one
kinematic variable usually called $q^2$ upon which the form factor
depends.  From energy-momentum conservation, $p_K = p_\pi + q_\ell + q_\nu$
where $p_K$ is the energy-momentum 4-vector of the initial state kaon,
$p_\pi$ refers to the final state pion, and $q_\ell$ and $q_\nu$,
refers to the final state lepton and neutrino.  The momentum transferred
to the leptons $q = q_\ell + q_\nu$, and we have already introduced
$q^2$.

\begin{figure}[tb]
\begin{center}
\includegraphics[width=0.7\textwidth]{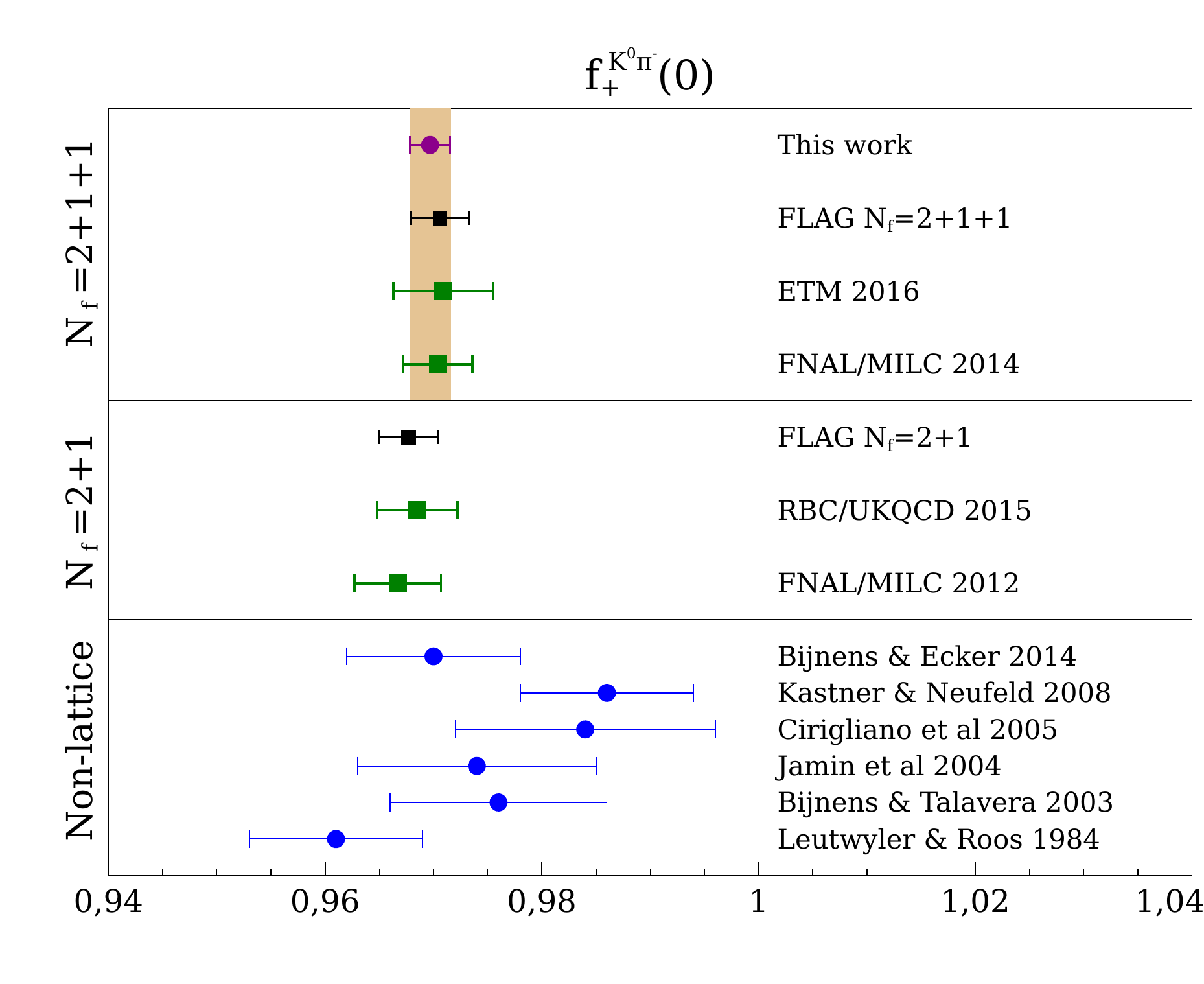}
\vspace{-5mm}
\caption{
\label{fig:MILC_Form_factors}
Comparison of calculations of the vector form factor  $f_+(q^2=0)$ with
$N_f=2+1+1$ and $2+1$ sea quark flavors, and other non-lattice QCD methods.  
From Ref.~\cite{Bazavov:2018kjg}.
}
\end{center}
\end{figure}

To determine $|V_{us}|$ from experiment, we could determine the vector
form factor $f_+(q^2)$ and use the differential decay rate of the
kaon; however, it is convenient to just calculate $f_+(q^2=0)$ and
use experimental input that determines $|V_{us}| f_+(0)$. 
We would like to start our story in 2014, when the experimental
value was $|V_{us}| f_+(0) = 0.2165(4)$.  At that time, FNAL/MILC had
an $N_f=2+1+1$ value $f_+(0)=0.9704(24)(22)$ where the first error was
statistical and the second systematic.  So, in 2014, the experimental
error was 0.18\%, but the theory error was 0.34\%.  Moving ahead to late 2018,
the experimental result had been slightly updated~\cite{Moulson:2017ive}
to:
$|V_{us}| f_+(0) = 0.21654(41)$.  Also FNAL/MILC ~\cite{Bazavov:2018kjg}
updated their calculation resulting in 
\begin{equation}
f_+(0) = 0.9696(15)_{\rm stat}(12)_{\rm sys} = 0.9696(19)\;,
\end{equation}
so the theory error is 0.20\%, quite comparable to the experimental error.
The FLAG 2019 average for $N_f=2+1+1$ is $f_+(0) = 0.9706(27)$.  The FLAG
result has a larger error because Ref.~\cite{Bazavov:2018kjg} had 
not been published before the FLAG deadline for inclusion in the latest
edition.  Figure~\ref{fig:MILC_Form_factors}, 
taken from Ref.~\cite{Bazavov:2018kjg}, depicts the
most relevant lattice calculations for $N_f=2+1$ and $2+1+1$.  It also contains results from a number of other theoretical approaches whose errors are
larger than those from lattice QCD.

\begin{figure}[tb]
\begin{center}
\includegraphics[width=0.65\textwidth]{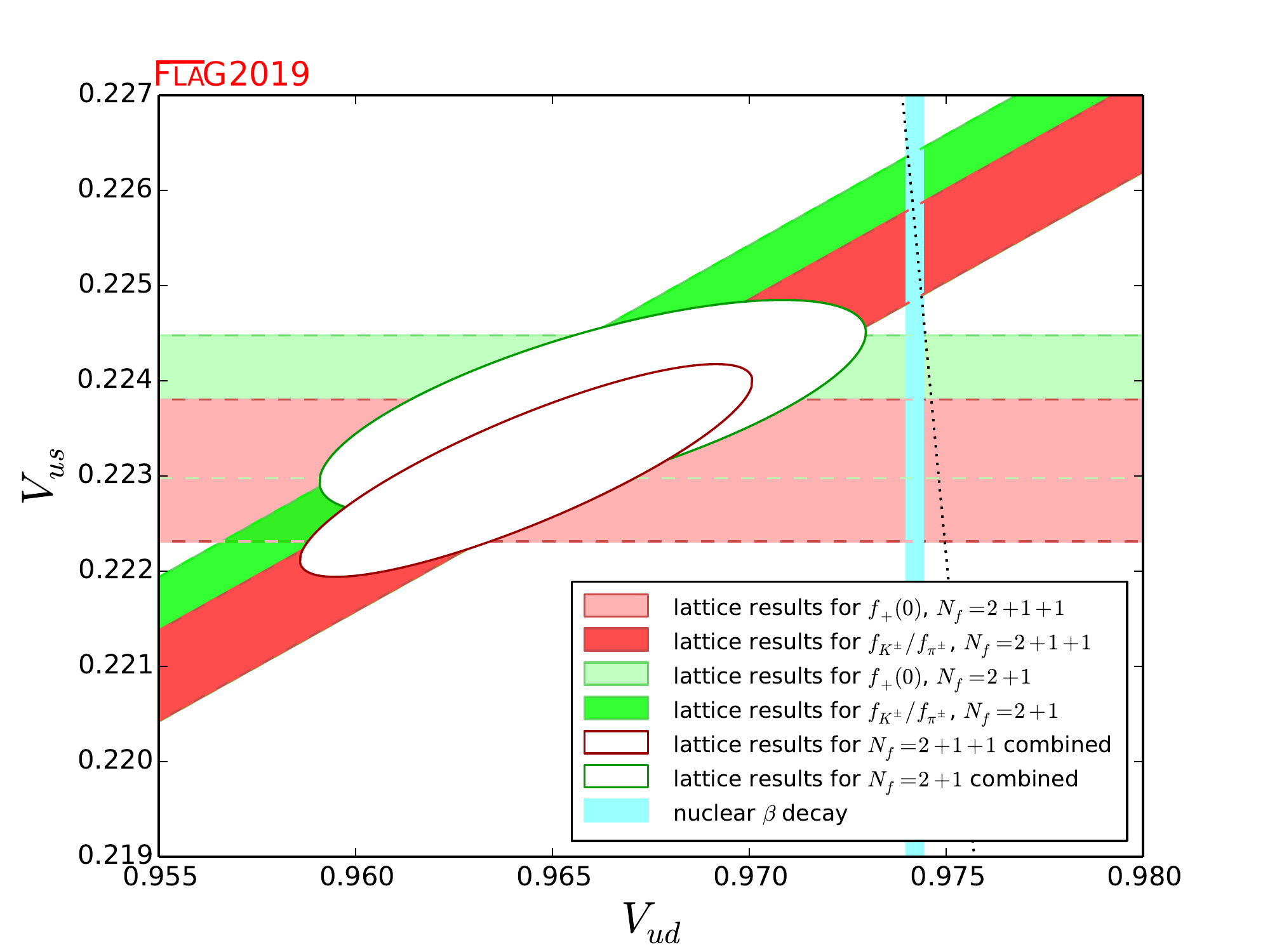}
\vspace{-5mm}
\caption{
\label{fig:FLAGVusVersusVud}
First row unitarity plot from FLAG~\cite{Aoki:2019cca}.
}
\end{center}
\end{figure}

\begin{figure}[h!bt]
\begin{center}
\includegraphics[width=0.55\textwidth]{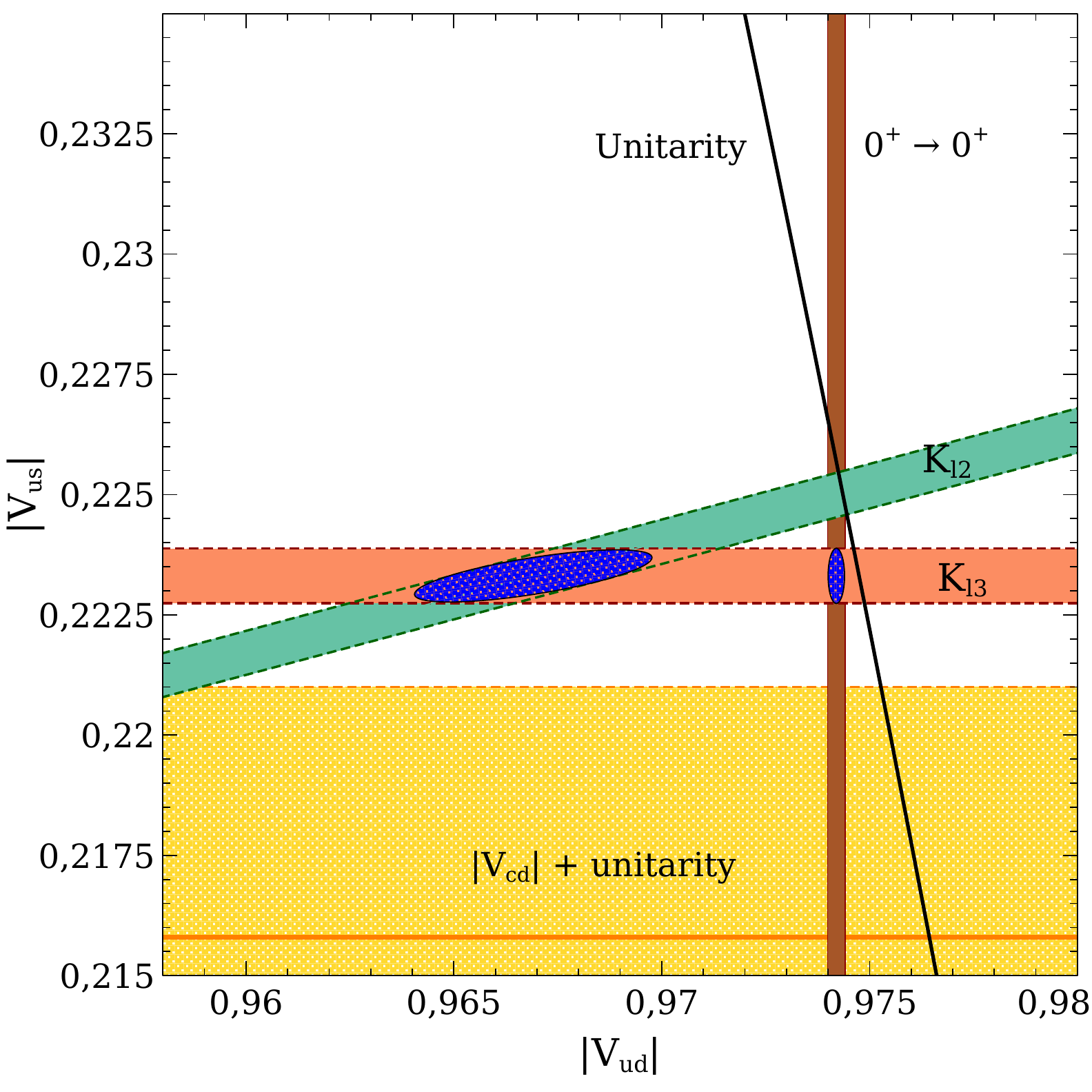}
\vspace{-5mm}
\caption{
\label{fig:FNALMILCfirstrow}
First row unitarity plot from FNAL/MILC~\cite{Bazavov:2018kjg}.
}
\end{center}
\end{figure}

We now turn to the $V_{ud}$-$V_{us}$ plane to consider the implications of
what we have just found.
The ratio of decay constant determines the ratio $|V_{us}/V_{ud}|$ and
thus an angled band in the plane of the two CKM matrix elements.  
The semileptonic kaon decay determines a horizontal band for $|V_{us}|$.
Nuclear $\beta$-decay provides a fairly precise value of $|V_{ud}|$, i.e.,
a vertical band.  In addition, because $|V_{ub}|$ is so small, unitarity
determines a very narrow arc of a circle in the plane.
Figure~\ref{fig:FLAGVusVersusVud} describes the situation according to 
FLAG~\cite{Aoki:2019cca}.
We see that the two white ellipses that come from the intersection of
leptonic and semileptonic decay bands show some tension with both
unitarity and $V_{ub}$ from $\beta$-decay.  This graph has results for both
$N_f=2+1$ and $2+1+1$ lattice QCD calculations.
It should be noted that the decay constant ratio is in reasonable
agreement with $\beta$-decay and unitarity.  In Fig.~\ref{fig:FNALMILCfirstrow},
we show the results from Ref.~\cite{Bazavov:2018kjg}.  
In this case, we have $N_f=2+1+1$ and
both blue ellipses show tension with unitarity.  The narrow horizontal band is
labeled $K_{\ell 3}$, a common notation for kaon semileptonic decay, while
the angled band is labeled $K_{\ell 2}$ a notation for kaon leptonic decay.
In addition, there is a wide horizontal band which is determined from the
known value of $|V_{cd}|$ and the assumption of CKM unitarity.  (Recall that
in the Wolfenstein representation, $V_{cd} = - V_{us}$.)


\begin{figure}[tb]
\begin{center}
\includegraphics[width=0.8\textwidth]{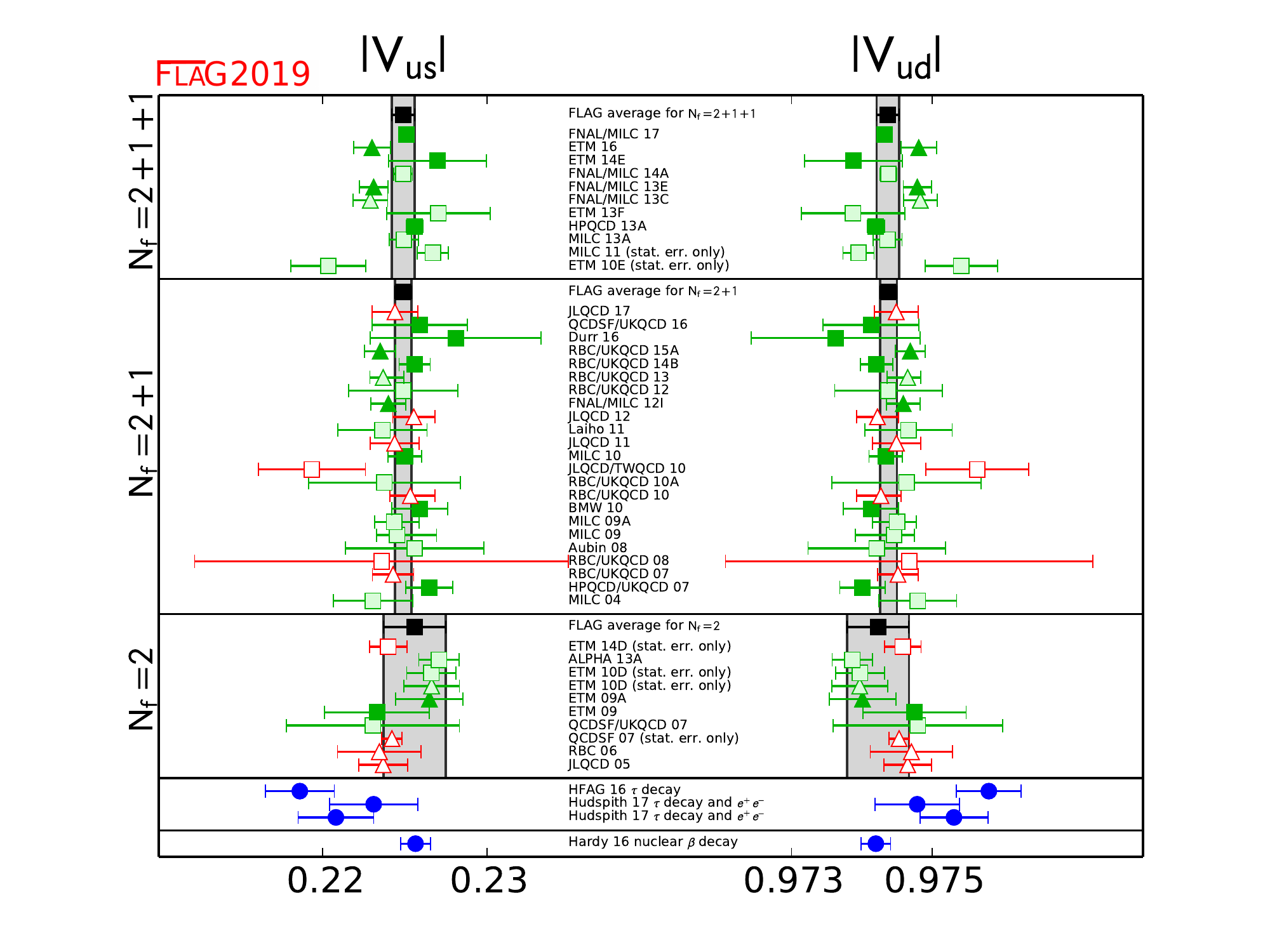}
\vspace{-5mm}
\caption{
\label{fig:FLAG_Vusud}
Determination of $|V_{us}|$ and $|V_{ud}|$ from leptonic and semileptonic
decays from FLAG~~\cite{Aoki:2019cca}.  Squares denote results from leptonic
decays and triangles are used for semileptonic decays.
}
\end{center}
\end{figure}

We have a summary of determination of $|V_{us}|$ and $|V_{ud}|$
from FLAG in Fig.~\ref{fig:FLAG_Vusud}.  This plot assumes unitarity to go from either band
to a value.  Squares denote values from leptonic decays, and triangles are used
for semileptonic decays.  The tension we saw in the previous two plots is
seen as a difference between triangles and squares, which is most noticeable
for $N_f=2+1+1$.  FLAG results for $N_f=2+1+1$ are:
\begin{eqnarray}
|V_{us}|&=0.2249(7)\\
|V_{ud}|&=0.97437(16)
\end{eqnarray}
The blue values near the bottom of the plot show results based on
standard model analysis of $\tau$ decay and nuclear $\beta$-decay.
Please see the FLAG report~\cite{Aoki:2019cca} for references.

\section{Second Row}
Returning to expression \ref{eq:ckmmatrix}, we find several semileptonic
decays on the row below $ \bf{ V_{cd} }$, and leptonic decays of $D$
and $D_s$ on the following row.  We will consider the leptonic decays first.

\begin{figure}[tb]
\begin{center}
\includegraphics[width=0.7\textwidth]{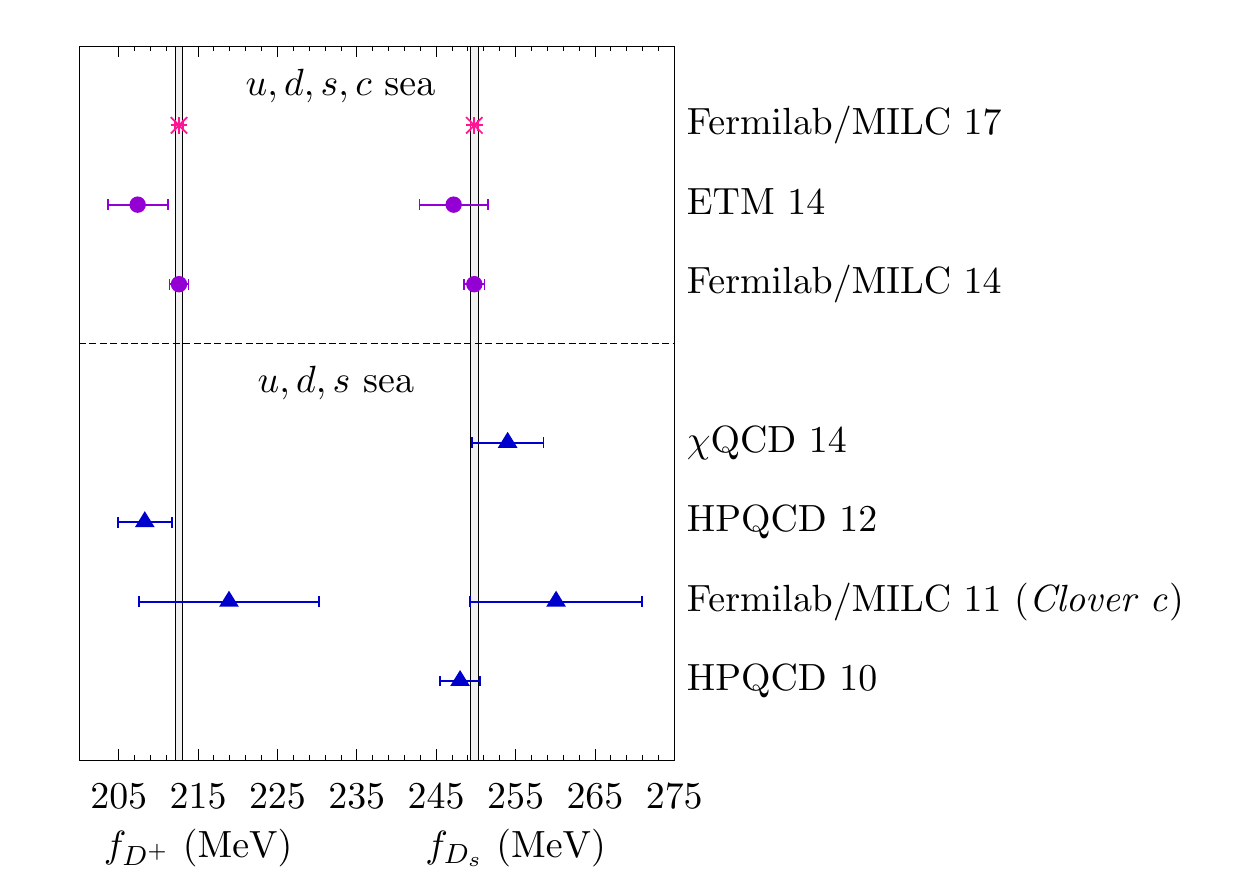}
\vspace{-5mm}
\caption{
\label{fig:fD_summary}
Comparison of recent calculations of $f_{D^+}$ and $f_{D_s}$ with $N+f=2+1+1$
and $2+1$ ~\cite{Bazavov:2017lyh}.
}
\end{center}
\end{figure}

It has been almost 15 years since the initial test of $D$
meson decay constants at CLEO-c~\cite{Artuso:2005ym}.
The unquenched calculations were based
on $N_f=2+1$ ensembles with asqtad quarks~\cite{Aubin:2005ar}.
Those early measurements
and calculations had errors of about 8\%.  By 2014, the theoretical error
had fallen by more than a factor of ten.  In 2017, using $N_f=2+1+1$
ensembles with HISQ quarks, errors on the charm decay constants have dropped
below $0.3$\%, considerably below the precision of results with $N_f=2+1$.
Figure~\ref{fig:fD_summary} from Ref.~\cite{Bazavov:2017lyh} depicts
several recent calculations of the $D_{(s)}$ meson decay constants.
The interested reader can find a more complete set of values in
Ref.~\cite{Aoki:2019cca}.  The results in Ref.~\cite{Bazavov:2017lyh} are:
\begin{equation}
f_{D^+} = 212.7(0.6) \mathrm{MeV}, \qquad f_{D_s} = 249.9(0.4) \mathrm{MeV}\;.
\end{equation}

The Particle Data Group~\cite{Tanabashi:2018oca} has compiled the results 
of several experiments to provide the best values for the product of decay
constant and CKM matrix elements.  The experimental values are:
$f_{D}|V_{cd}|=45.91(1.05) \mathrm{MeV}$ and
$f_{D_s}|V_{cs}|=250.9(4.0) \mathrm{MeV}$.  
These values have errors of 1.6--2.3\%, so the experimental error is going
to be dominant in determination of the two CKM matrix elements.  From
Ref.~\cite{Bazavov:2017lyh}, we have 
\begin{eqnarray}
|V_{cd}|_{{\rm SM},\, f_D}  &= 0.2152(5)_{f_D}(49)_{\rm expt}(6)_{\rm EM},\\
    |V_{cs}|_{{\rm SM},\, f_{D_s}} &= 1.001(2)_{f_{D_s}}(16)_{\rm expt}(3)_{\rm EM}\;,
\end{eqnarray}
where the errors are from lattice decay constant, experiment, and a structure
dependent electromagnetic correction. These values differ slightly from
those in FLAG~\cite{Aoki:2019cca}.  Earlier this year, some new results
for $D_s$ decay were published by BESIII~\cite{Ablikim:2018jun}.  It was
found that $f_{D_s}|V_{cs}|=246.2(5.0) \mathrm{MeV}$.  This will bring
the world average down and help improve the second row unitary sum.
In fact, after Lattice 2019, the Heavy Flavor Averaging Group 
(HFLAV))~\cite{Amhis:2019ckw} updated its results for many quantities 
of interest including the leptonic decays of $D$ and $D_s$ mesons.
Their current world averages are
$f_{D}|V_{cd}|=46.1(1.1) \mathrm{MeV}$ and
$f_{D_s}|V_{cs}|=247.8(3.1) \mathrm{MeV}$.  These errors are 2.4 and 1.3\%,
respectively, which is an improvement for $D_s$. HFLAV uses the FLAG results
for the decay constants, which have slightly more generous errors than
those in Ref.~\cite{Bazavov:2017lyh}.  The FLAG value of $f_{D}$ is also 
slightly smaller.  
With that input, HFLAV finds $|V_{cd}| = 0.2173(51)_{expt}(7)_{LQCD}$ and 
$|V_{cs}| = 0.991(13)_{expt}(2)_{LQCD}$.

\begin{figure}[tb]
\begin{center}
\includegraphics[width=0.7\textwidth]{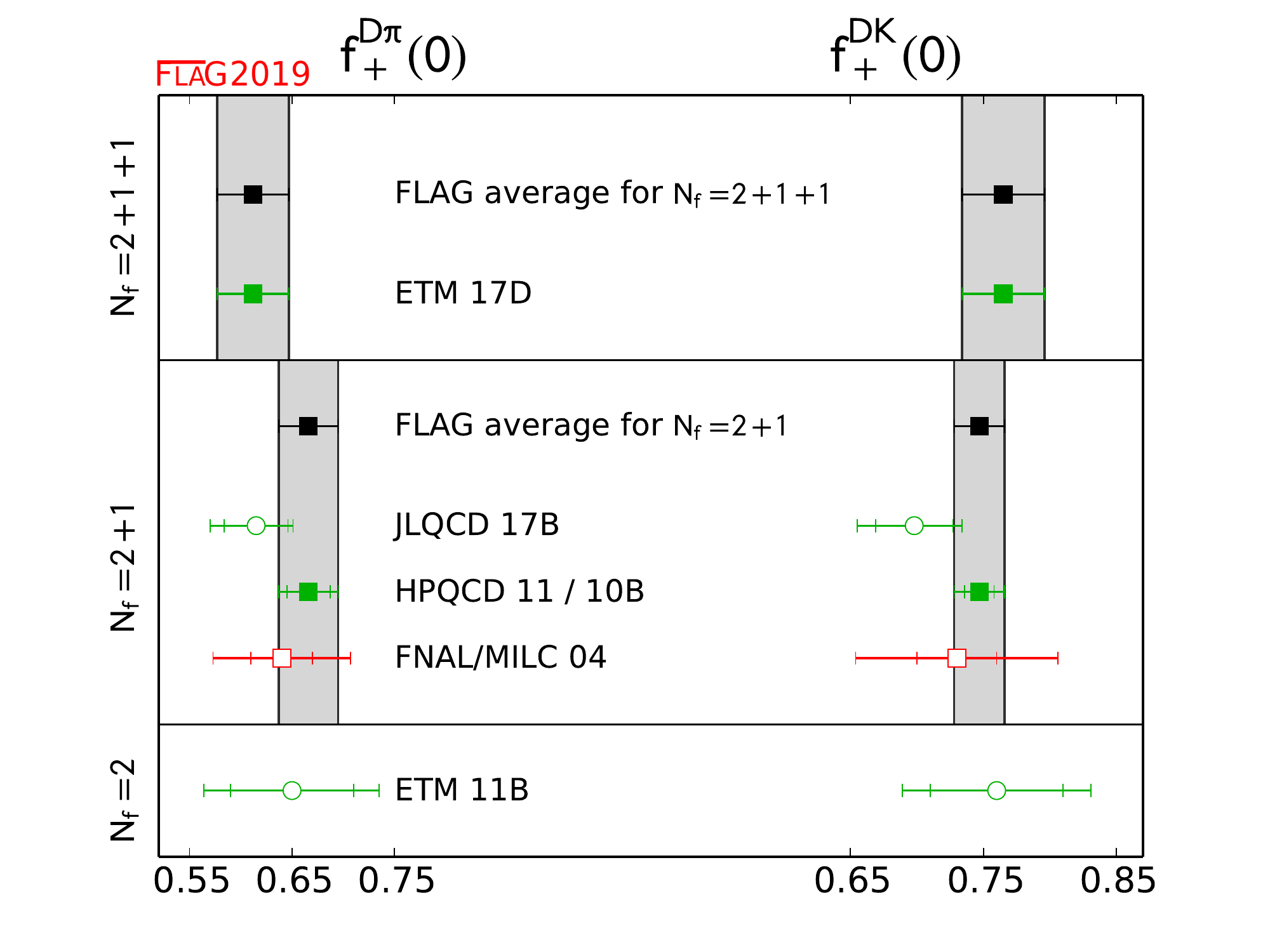}
\vspace{-5mm}
\caption{
\label{fig:DtoPiandDtoK}
FLAG~\cite{Aoki:2019cca} summary of results for $D$ semileptonic decays to $\pi$ or $K$.
}
\end{center}
\end{figure}

\begin{figure}[tb]
\begin{center}
\includegraphics[width=0.7\textwidth]{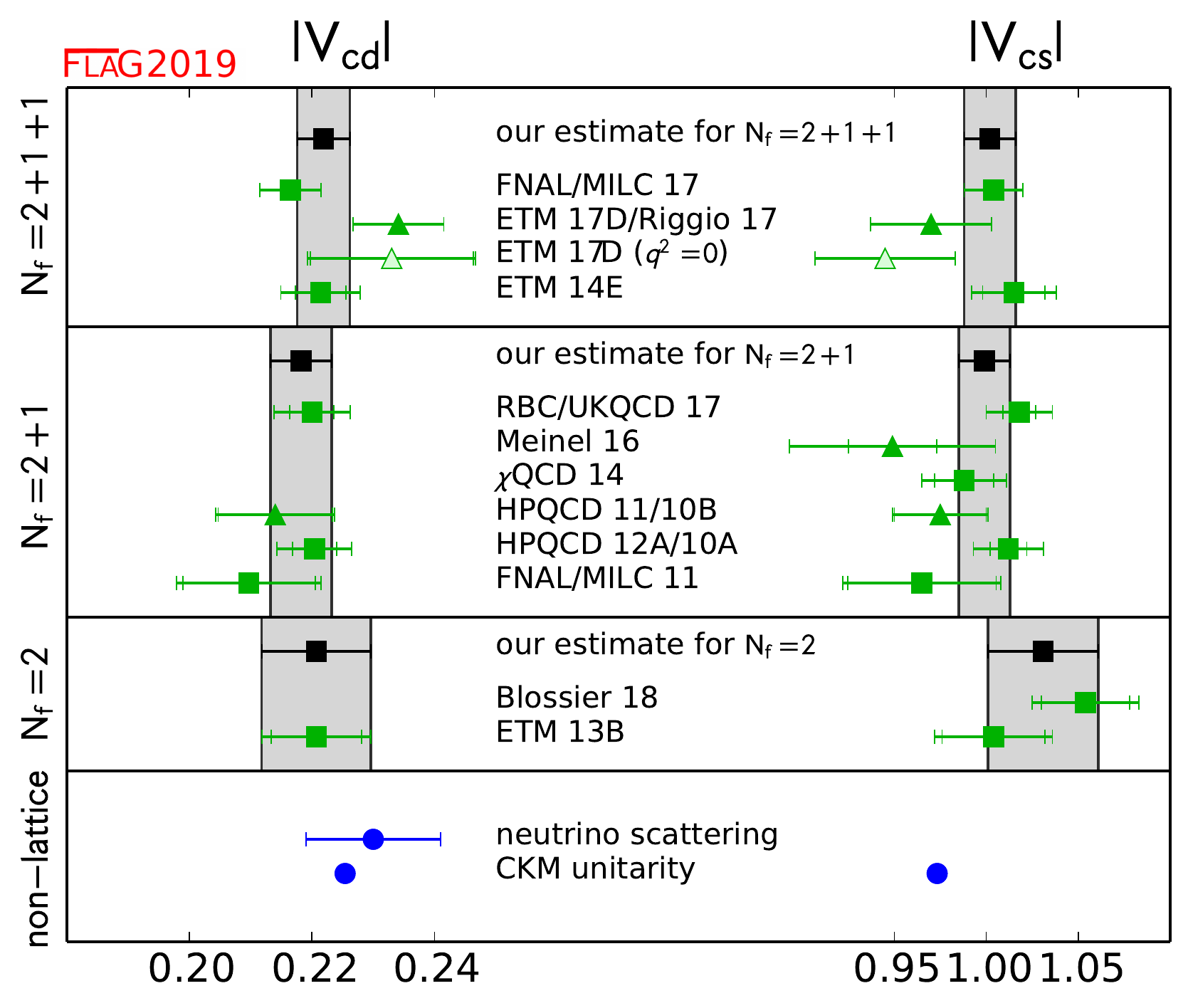}
\vspace{-5mm}
\caption{
FLAG~\cite{Aoki:2019cca} summary of results for $|V_{cd}|$ and $|V_{cs}|$.
Squares are used for results from
leptonic decays, and triangles for semileptonic decays.
\label{fig:VcdandVcs}
}
\end{center}
\end{figure}

For charm semileptonic decays, a single form factor describes the major
contribution to the decay rate.  The analysis is often restricted to
$q^2=0$.  In that case, we can use the world average values provided by
the Heavy Flavor Averaging Group (HFLAV)~\cite{Amhis:2019ckw}:
$f_+^{D\pi}(0)|V_{cd}|= 0.1426(18)$ and $f_+^{DK}(0)|V_{cs}|= 0.7180(33)$.
These values have been updated since Lattice 2019 and do not agree with
the values quoted in my slides on the conference indico site.  
In particular, the second value has
decreased.  The experimental errors are 1.3\% and 0.5\% for the two decays.
Figure~\ref{fig:DtoPiandDtoK} shows the FLAG averages for both form factors at
$q^2=0$.  For both $N_f=2+1$ and $2+1+$ there is only a single result.
Consider the form factors for $D\to K \ell\nu$, as that has the smaller
experimental error,
the $N_f=2+1+1$ result from ETM ~\cite{Lubicz:2017syv}
is $0.765(31)$, i.e., a error of 4.2\%.
The $N_f=2+1$ result from HPQCD~\cite{Na:2010uf,Na:2011mc}
is 0.747(19) or a 2.5\% result.
We see that the theoretical errors need to decrease by a factor of 2--4 to
match the current experimental precision.  It should be noted that ETM has
form factor results as a function of $q^2$.  Also, FNAL/MILC are expecting to
have a result with an error of about 2.1\%.  FLAG had summarized
the results for $|V_{cd}|$ and $|V_{cs}|$ which are displayed in
Fig.~\ref{fig:VcdandVcs}.  As before, squares are used for results from
leptonic decays, and triangles for semileptonic decays.  These FLAG results
use an older value of  $f_+^{DK}(0)|V_{cs}|= 0.7226(34)$.
The result labeled Meinel 16 is from baryon decay, and the result labeled
ETM 17D/Riggio 17 uses non-zero values of $q^2$ to extract $|V_{cs}|$ from
the differential decay rate.  Experimental errors dominate for
the leptonic decays, while the theory error is dominant for semileptonic
decays.  Here are some key FLAG results.


\begin{eqnarray}
{\rm leptonic~decays}, N_f=2+1+1:&|V_{cd}| = 0.2166(7)(50)\,, &|V_{cs}| = 1.004  (2)(16) \,, \\
{\rm leptonic~decays}, N_f=2+1:  &|V_{cd}| = 0.2197(25)(50)\,, &|V_{cs}| = 1.012  (7)(16) \,, \\
{\rm semileptonic~decays}, N_f=2+1+1:&|V_{cd}| = 0.2341(74)\,, &|V_{cs}| = 0.970  (33) \,, \\
{\rm semileptonic~decays}, N_f=2+1:  &|V_{cd}| = 0.2141(93)(29)\,, &|V_{cs}| = 0.967  (25)(5) \,, \\
{\rm semileptonic~\Lambda_c~decay}, N_f=2+1:  &                    &|V_{cs}| = 0.949  (24)(51) \,, \\
{\rm FLAG2019}, N_f=2+1+1:&|V_{cd}| = 0.2219(43)\,, &|V_{cs}| = 1.002  (14) \,, \\
{\rm FLAG2019}, N_f=2+1:&|V_{cd}| = 0.2182(50)\,, &|V_{cs}| = 0.999 (14) 
\end{eqnarray}

Let's consider second row unitarity.  
Using their most recent leptonic decay results~\cite{Bazavov:2017lyh} 
Fermilab/MILC have:
$|V_{cd}|^2 + |V_{cs}|^2 + |V_{cb}|^2
 = 1.049(2)_{|V_{cd}|}(32)_{|V_{cs}|}(0)_{|V_{cb}|}$,
which is just over 1.5 $\sigma$ from 1.  The FLAG 2019 value for $N_f=2+1$ and
2+1+1 is 1.05(3), which is similar.  
The unitarity sum from ETM~\cite{Riggio:2017zwh}, which is based
on semileptonic decays, is 0.996(64). It agrees very well with unitarity,
but with a larger error than what is seen from leptonic decays.  We had
already seen that semileptonic decays favor a smaller value of $|V_{cs}|$ than
leptonic decays do, so the improved agreement with unitarity should not
be a surprise.  If we consider the latest HFLAV update, we have:
$|V_{cd}|^2 + |V_{cs}|^2 + |V_{cb}|^2
 = 1.029(2)_{|V_{cd}|}(26)_{|V_{cs}|}(0)_{|V_{cb}|}$,
which is only about 1.1 $\sigma$ away from 1.  Thus, we have reasonable
agreement with unitarity for both leptonic and semileptonic charm meson
decays, with a few percent accuracy.  It will, or course, be interesting 
to increase the precision of this test in the future.

\section{$B$ Hadron Decays}
Leptonic and semileptonic decays of hadrons containing a $b$ quark have
been studied using lattice QCD.  Mesonic decays have been extensively studied.
Recently, Meinel and his collaborators have been looking at various baryon
semileptonic decay form factors~\cite{Detmold:2015aaa}.  
In addition to tree level decays such as
$B\to\pi\ell\nu$, which can be used to determine $|V_{ub}|$, there have been
studies of rare decays that involve flavor changing 
neutral currents~\cite{Du:2015tda}.
Such decays vanish at tree level in the Standard Model, so they are a good
place to search for new physics.  They provide a complementary method to
$B$ mixing studies for determining $|V_{td}|$ and $|V_{ts}|$.  There have
been some interesting tensions between recent Standard Model predictions
and LHCb measurements, including hints of lepton universality violation.
(See Ref.~\cite{Du:2015tda} for a summary.) 
Although these issues are very interesting, we will need to stay focused
on determination of CKM matrix elements.

\begin{figure}[tb]
\begin{center}
\includegraphics[width=0.7\textwidth]{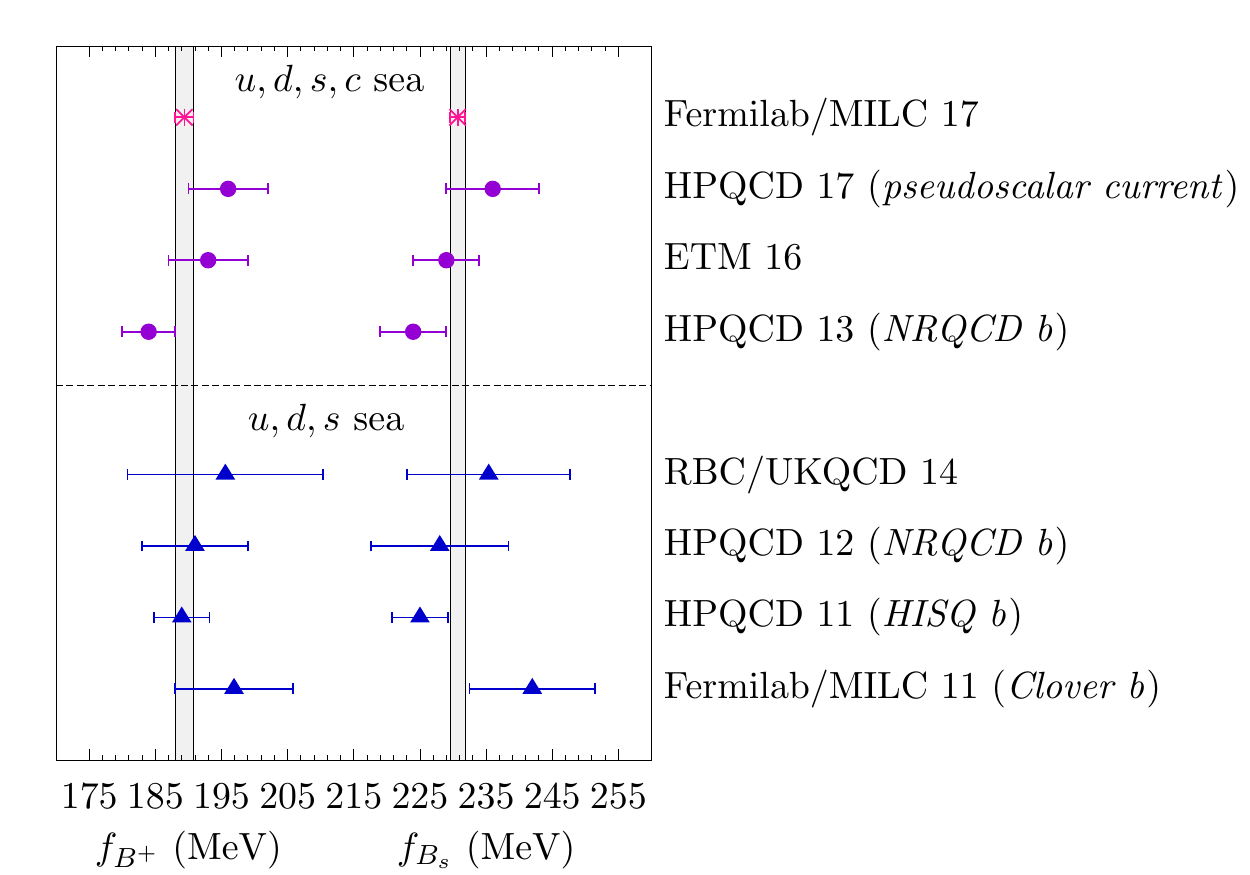}
\vspace{-5mm}
\caption{Comparison of recent results for $f_{B^+}$ and $f_{B_s}$ from 
Ref.~\cite{Bazavov:2017lyh}.
\label{fig:fB_summary}
}
\end{center}
\end{figure}

In Fig.~\ref{fig:fB_summary}, we show a compilation of results from
Ref.~\cite{Bazavov:2017lyh}.  The latest results from FNAL/MILC have errors
under 1.3 MeV, i.e., $< 0.7\%$.  These results are in good agreement with
prior results that had errors as small as 5--7 MeV.  Unfortunately,
the BaBar and Belle experimental measurements of $B$ decay
do not agree very well and have large errors, so the determination of
$|V_{ub}|$ is very much limited by the required experimental input.  In the
next few years, Belle II should provide much higher precision results.
Additional details and historical calculations may be found in 
Refs.~\cite{Aoki:2016frl} and \cite{Aoki:2019cca}.

As mentioned above, $B$ semileptonic and rare decays have been studied both
experimentally and theoretically.  Several decays such a
$B\to \pi\ell\nu$, $B_s\to K^{(*)} \ell\nu$, and $\Lambda_b\to p \ell \nu$
all depend on $|V_{ub}|$.  These decays can be used to determine
$|V_{cb}|$: $B\to D^{(*)} \ell\nu$,  $B_s\to D_s^{(*)} \ell\nu$, and
$\Lambda_b\to \Lambda_c \ell \nu$.  Recall that when a particular CKM
matrix element can be determined from multiple decays, if the standard model
predictions of the decay rates do not imply identical values of the
particular CKM matrix element, that would be evidence for new physics not
in the standard model.  In addition, since $\ell$ can be an electron, muon or
$\tau$, there is an opportunity to test lepton universality.  Recently,
some tensions with lepton universality have been seen, and this is a
fertile area of study.  Unfortunately, we do not have time to discuss
this in detail.  Interesting rare decays include $B^0\to \mu^+\mu^-$,
$B_s\to \mu^=\mu^-$, and $B\to K \ell^=\ell^-$.

\begin{figure}[tb]
\begin{center}
\includegraphics[width=0.7\textwidth]{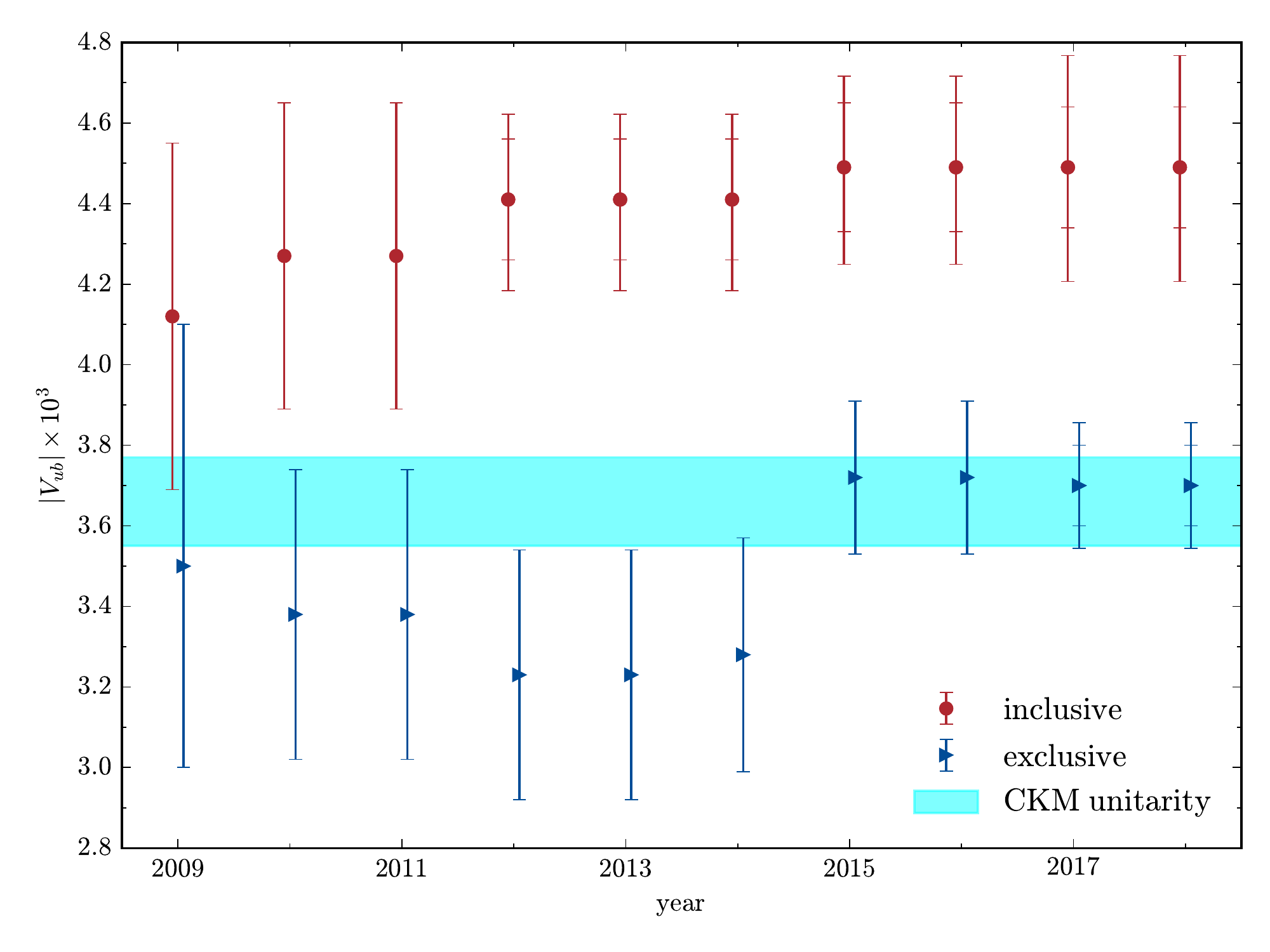}
\vspace{-5mm}
\caption{History of $|V_{ub}|$ as determined from both inclusive and
exclusive measurements.  
\label{fig:Vub}
}
\end{center}
\end{figure}

There has been a long standing difference between the values of $|V_{ub}|$
and $|V_{cb}|$ as determined from exclusive and inclusive decay results.
The history of the comparison from 2009 to 2018 is shown in Fig.~\ref{fig:Vub}.
When the precision of each determination was low in 2009, there was not a
tension between the two values; however, by 2014 the difference had
grown.  In 2015, the difference was reduced, but there has not been much
change since then.  The plot also has a horizontal band corresponding to CKM
unitarity.

\begin{figure}[tb]
\begin{center}
\includegraphics[width=0.7\textwidth]{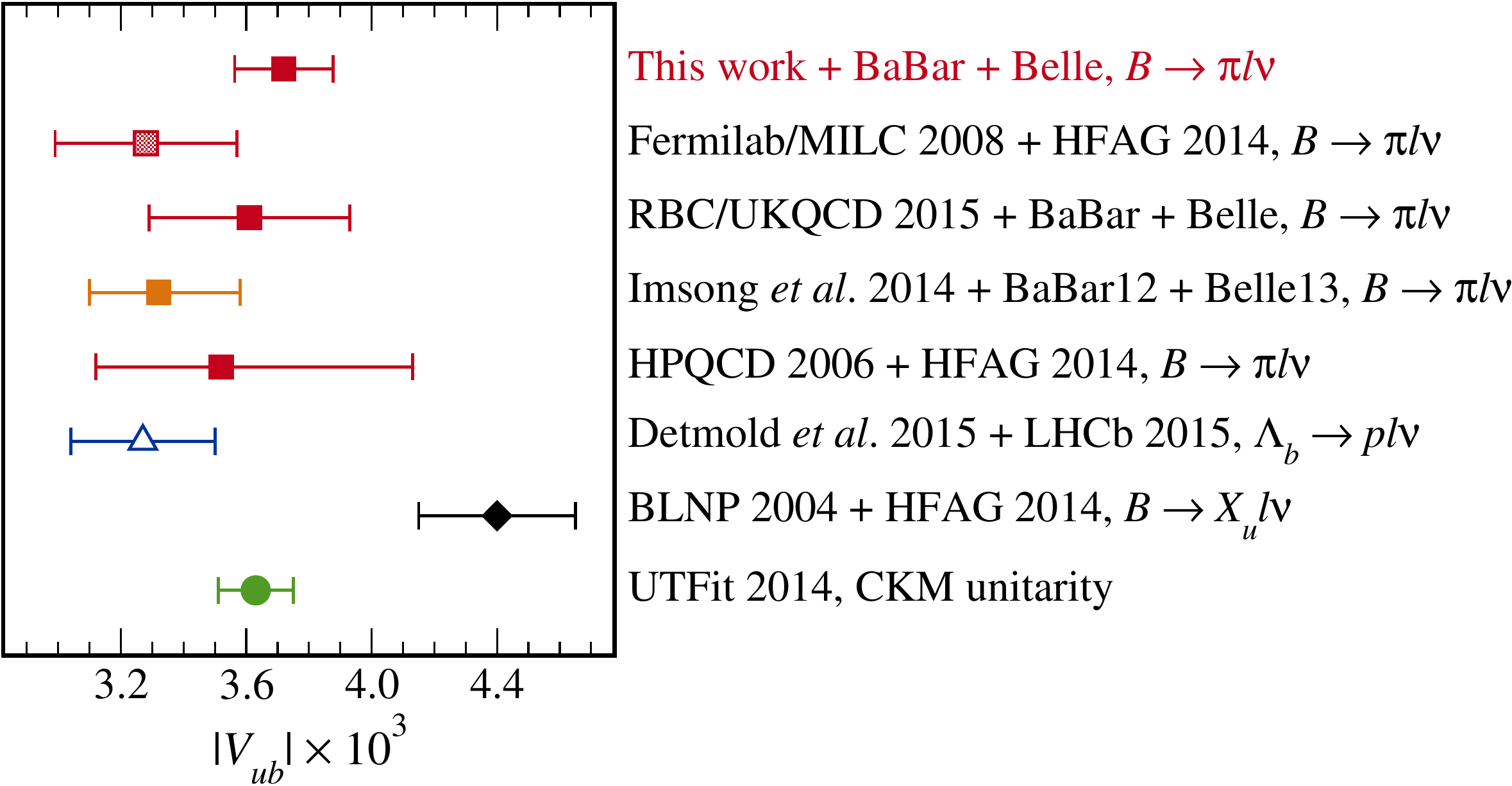}
\vspace{-5mm}
\caption{Comparison of several determinations of $|V_{ub}|$ from
Ref.~\cite{Lattice:2015tia}.
\label{fig:Vub_compare}
}
\end{center}
\end{figure}

\begin{figure}[tb]
\begin{center}
\includegraphics[width=0.7\textwidth]{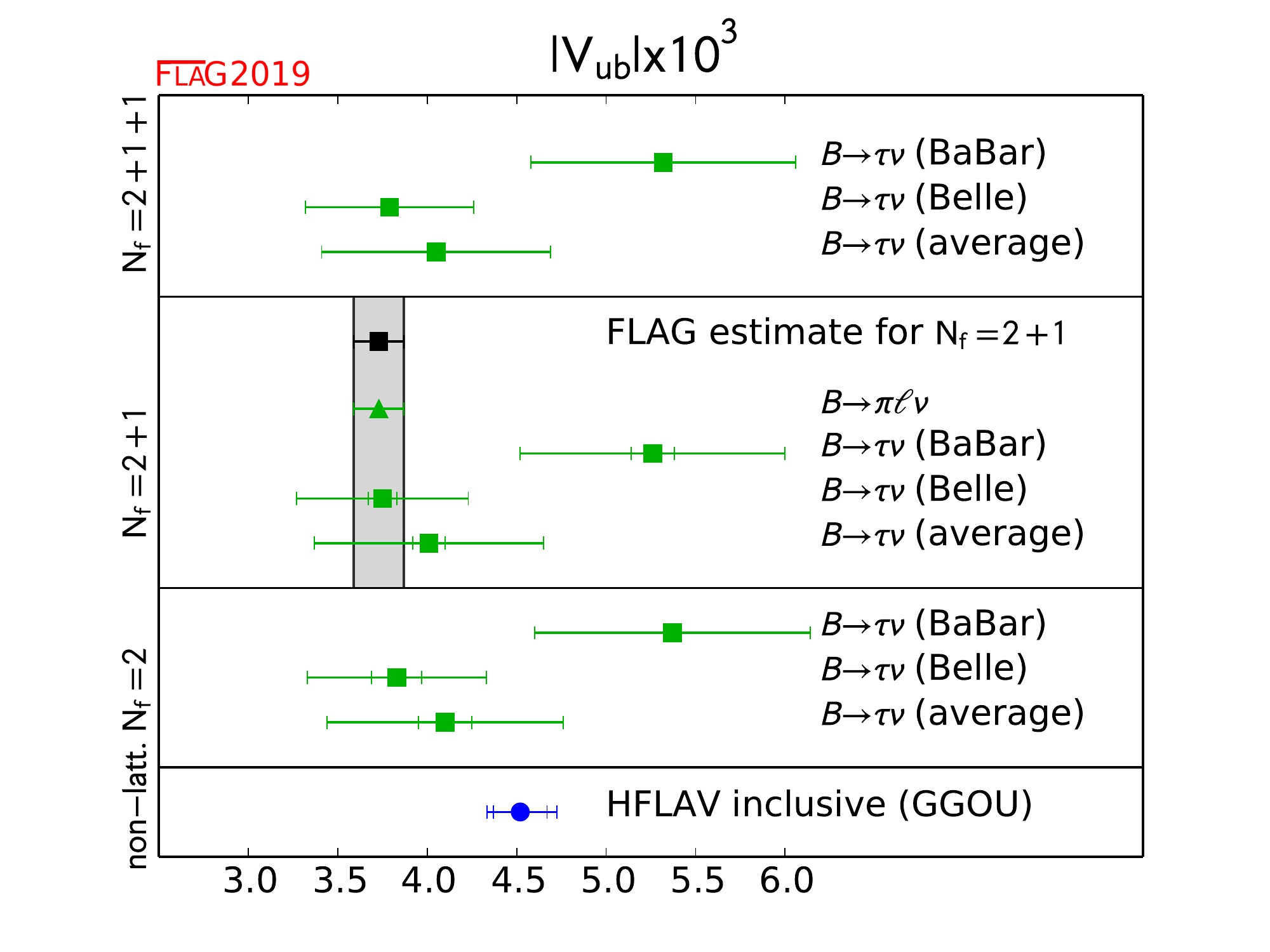}
\vspace{-5mm}
\caption{FLAG summary plot for $|V_{ub}|$ from Ref.~\cite{Aoki:2019cca}.
\label{fig:FLAG_Vub}
}
\end{center}
\end{figure}

The change in the value of $|V_{ub}|$ as determined from exclusive decays
resulted from a new form factor calculation from the Fermilab Lattice and MILC
collaborations~\cite{Lattice:2015tia}.  Experimental results from both
BaBar and Belle were used to determine $|V_{ub}| = 3.72(16)\times 10^{-3}$.
Figure~\ref{fig:Vub_compare} from Ref.~\cite{Lattice:2015tia} shows both
their calculation (denoted ``This work'') and other results.  Their result
was in reasonable agreement with other exclusive
decay calculations, but with smaller error.  However, the black diamond labeled
BLNP, which comes from inclusive decays, is larger.
There have been no more recent published results for the $B\to\pi\ell\nu$
form factors.  The FLAG report summarizes form factor results and their
comparison with experiment (which determines $|V_{ub}|$).  We will not
reproduce those graphs here, but Fig.~\ref{fig:FLAG_Vub}
shows the FLAG summary plot for
$|V_{ub}|$, which considers both leptonic and semileptonic decays,
and we quote the average value $3.73(14)\times 10^{-3}$, which
is a 3.8\% error.  The BaBar and Belle leptonic decay results do not agree
very well, so FLAG reports results based on each experiment and their
average.  The experimental error is dominant for all values of $N_f$.
Only for $N_f=2+1$ are there semileptonic form factors from lattice QCD.
The lattice QCD form factor and experimental measurements from BaBar
and Belle are jointly fit to a $z$-expansion with $|V_{ub}|$ as
a free parameter.  One sees that currently the error from semileptonic
determination of $|V_{ub}|$ is smaller than the determination from 
leptonic decays.

An interesting story has emerged over the last several years regarding
$V_{cb}$.  In this case, both semileptonic decays $B\to D\ell\nu$
and $B\to D^*\ell\nu$ have been measured.  The inclusive measurements also
have been used to determine $|V_{cb}|$.  In 2014, Ref.~\cite{Bailey:2014tva}
updated the form factor for $B\to D^*$ at zero-recoil, and used the
experimental average from HFLAV~\cite{Amhis:2012bh}
(then using a different acronym) to obtain 
$|V_{cb}|=(39.04\pm 0.49_{expt} \pm 0.49_{QCD} \pm 0.19_{QED} )\times 10^{-3}.$
The inclusive value from Gambino and Schwanda~\cite{Gambino:2013rza}, was
$(42.42\pm 0.86)\times 10^{-3}$, which differs by $3.0\sigma$ from the
exclusive value.  The next year, results for the $B\to D$ semileptonic decay
became available~\cite{Lattice:2015rga},
which gave a slightly larger value of $|V_{cb}|$, but with larger errors.
Updated results from Belle became available for $B\to D$ which
Bigi and Gambino~\cite{Bigi:2016mdz} analyzed using the BGL 
form factor parametrization~\cite{Boyd:1997kz}.  They found
$|V_{cb}| = (40.49\pm 0.97)\times 10^{-3}$.  The next year, two groups
looked at new Belle data~\cite{Abdesselam:2017kjf} for $B\to D^*$.
Both Bigi, Gambino, Schacht~\cite{Bigi:2017njr};
and Grinstein and Kobach~\cite{Grinstein:2017nlq} found about a 10\%
difference when switching between CLN~\cite{Caprini:1997mu} and BGL 
parametrizations of the form factors.  At the time of FPCP 2018,
I thought the so-called $V_{cb}$ puzzle was largely 
resolved~\cite{Gottlieb:2018dja} as the exclusive value of $|V_{cb}|$ 
using the BGL parametrization was quite compatible with the inclusive
value, and the difference between CLN parametrized value 
and the inclusive value was only $1.36\sigma$.  However, the situation 
subsequently became more murky.  At first, Belle~\cite{Abdesselam:2018nnh}
put out a preprint in September, 2018, that supported the notion that
there is a 10\% difference when switching between CLN and BGL.  Then,
in April, 2019, version 3 of that preprint found that CLN and BGL 
parametrizations are quite compatible.  A month earlier, an analysis 
from BaBar~\cite{Dey:2019bgc}, using an unbinned fit, an angular
analysis, the BGL parametrization found
 $|V_{cb}| = (38.36\pm 0.90)\times 10^{-3}$.  This result is in good agreement
with previous values of  $|V_{cb}|$ based on exclusive decays.

\begin{figure}[!h]
\begin{center}
\begin{minipage}{0.49\textwidth}
\includegraphics[width=1\linewidth]{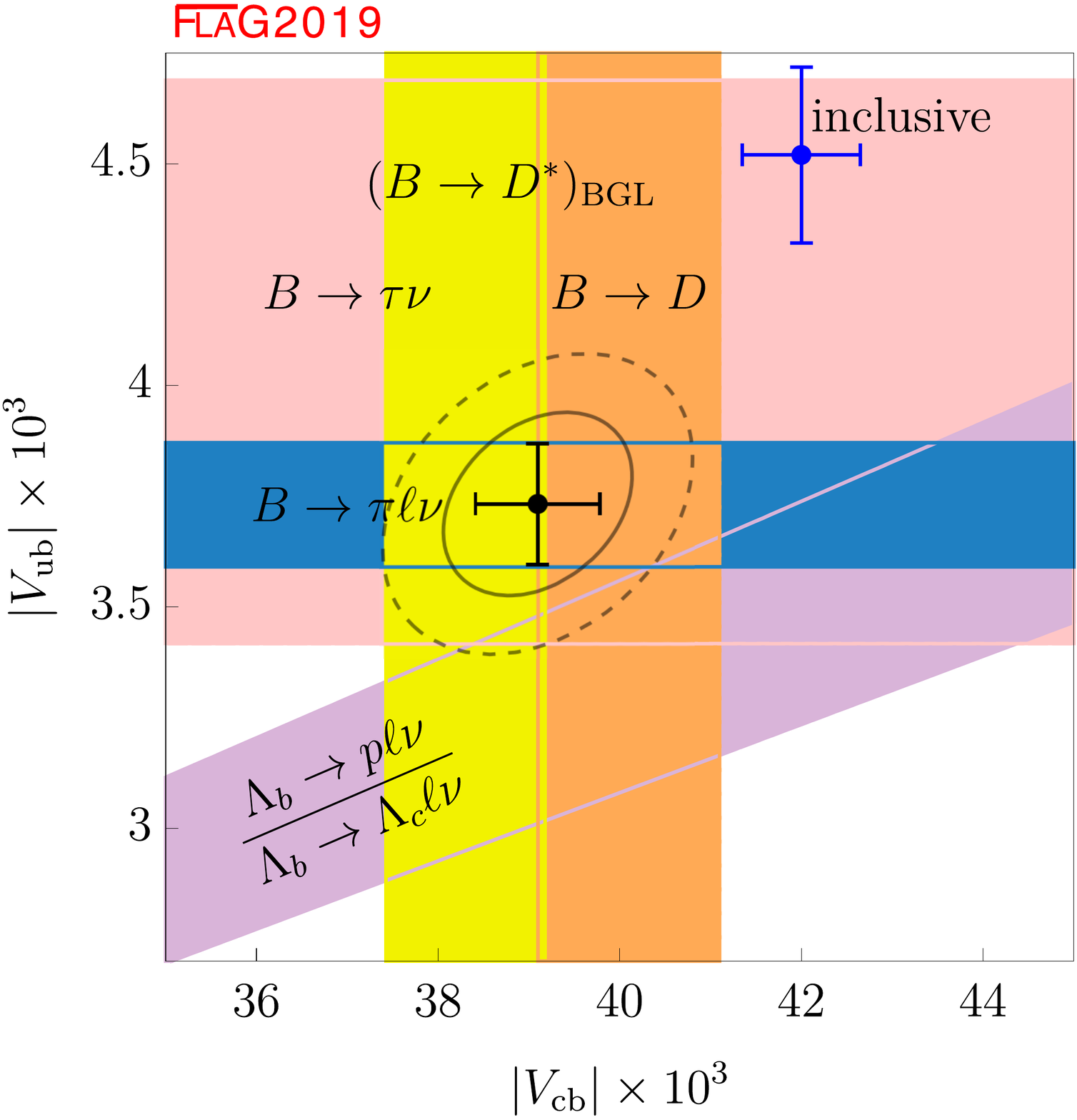}
\end{minipage}
\begin{minipage}{0.49\textwidth}
\includegraphics[width=1\linewidth]{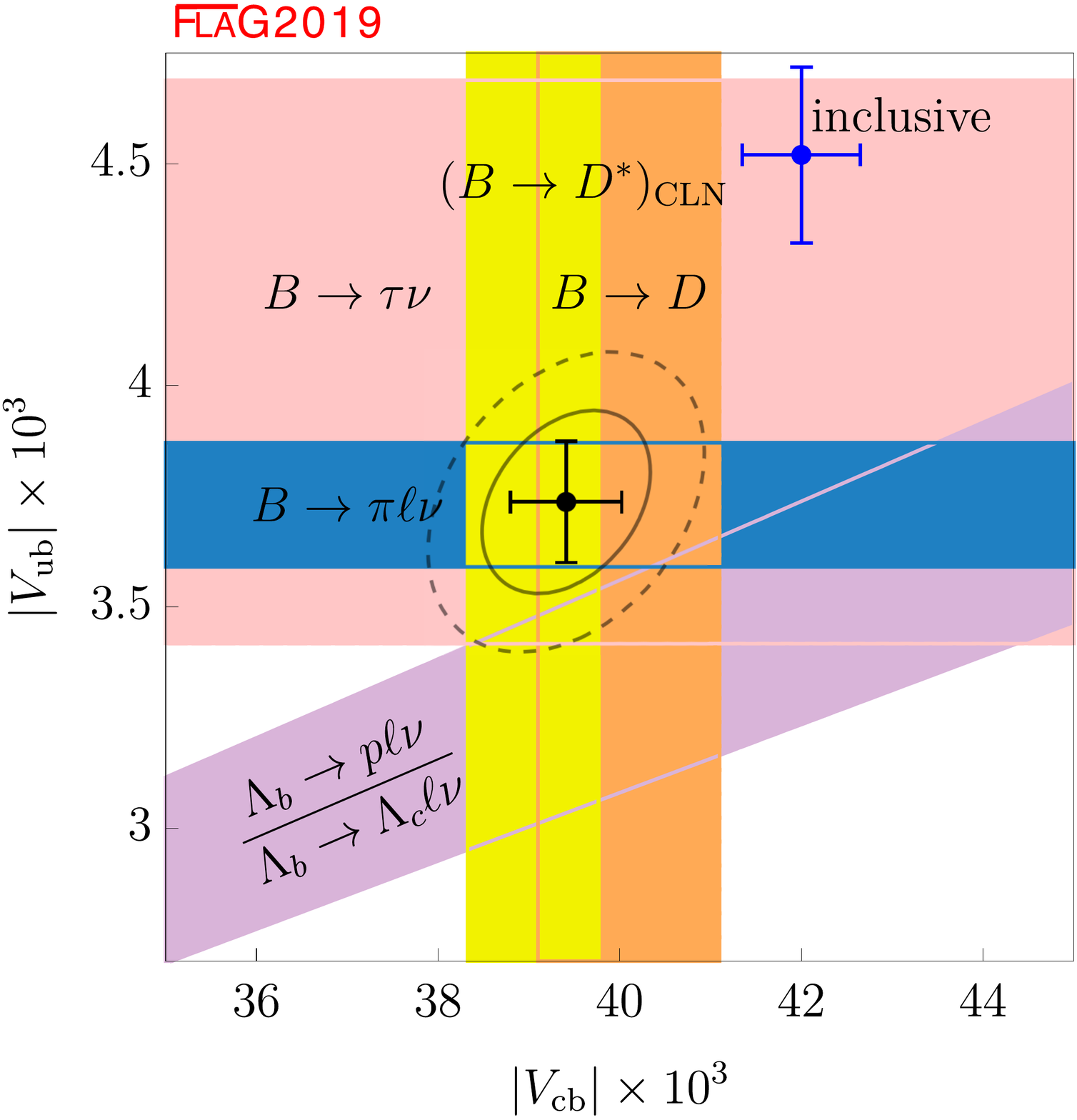}
\end{minipage}
\vspace{-2mm}
\caption{FLAG has summarized recent results for $|V_{ub}|$ and $|V_{cb}|$
\cite{Aoki:2019cca}.
On the left, the BGL parametrization for 
the $B\to D^*$ form factor is used.  On the right, CLN is used.
Further details may be found in Ref.~\cite{Aoki:2019cca}.
}
\label{fig:VubVcb}
\end{center}
\end{figure}

More details can be found in Andrew Lytle's review in these proceedings,
and in a recent paper by Gambino, Jung, and Schacht~\cite{Gambino:2019sif}.
After performing a number of fits to Belle data, the latter find
about a $2 \sigma$ difference between exclusive and inclusive
values for $|V_{cb}|$.  Thus, the puzzle remains until there are 
improvements in the lattice QCD calculations and the experiments.
The form factor calculations will likely be improved by four competing
groups: FNAL/MILC, HPQCD, JLQCD, and LANL/SWME.  The interested
reader should check the proceedings for their contributions.
Figure~\ref{fig:VubVcb} displays current results using either BGL or CLN
parametrizations.

\section{Third Row}
\begin{figure}[tb]
\begin{center}
\includegraphics[width=0.7\textwidth]{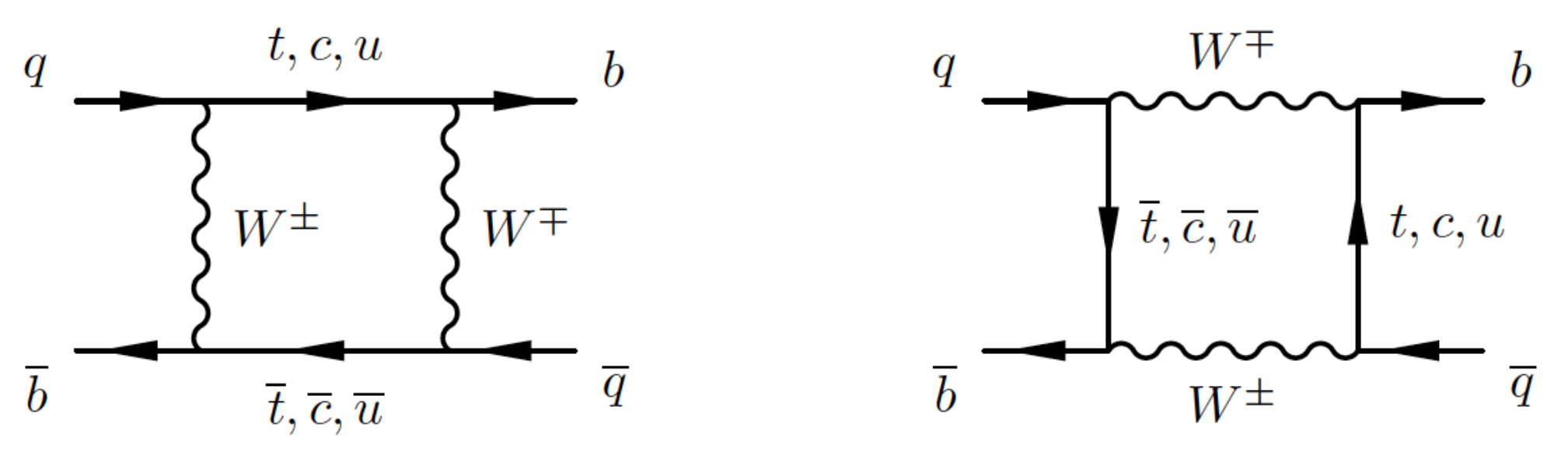}
\vspace{-5mm}
\caption{Box diagrams that contribute to neutral $B$ meson mixing from
Ref.~\cite{Bazavov:2016nty}.
\label{fig:box_diagrams}
}
\end{center}
\end{figure}

Neutral B meson mixing is a loop level process that
depends upon $V_{td}$ or $V_{ts}$ depending on whether we consider
$B$ or $B_s$ meson mixing.  
Figure\ref{fig:box_diagrams} from a recent paper by the Fermilab Lattice and
MILC Collaborations
\cite{Bazavov:2016nty} shows the
1-loop diagrams responsible for neutral $B$ meson mixing in the standard model.
The line labeled $q$ can either be a $d$ or $s$ quark.  Both the mass
difference and lifetime difference of the two resulting eigenstates are
measured in experiments.  A $CP$ violating phase is also determined.
A short distance expansion of the loops results in an effective weak
Hamiltonian involving 4-quark operators.  Because of the GIM and loop 
suppression of the mixing, this is a good place to look for BSM 
physics, which results in either operators that do not appear
in the standard model effective Hamiltonian, or modification of the coefficient
of the standard model operator.

\begin{figure}[tb]
\begin{center}
\includegraphics[width=0.7\textwidth]{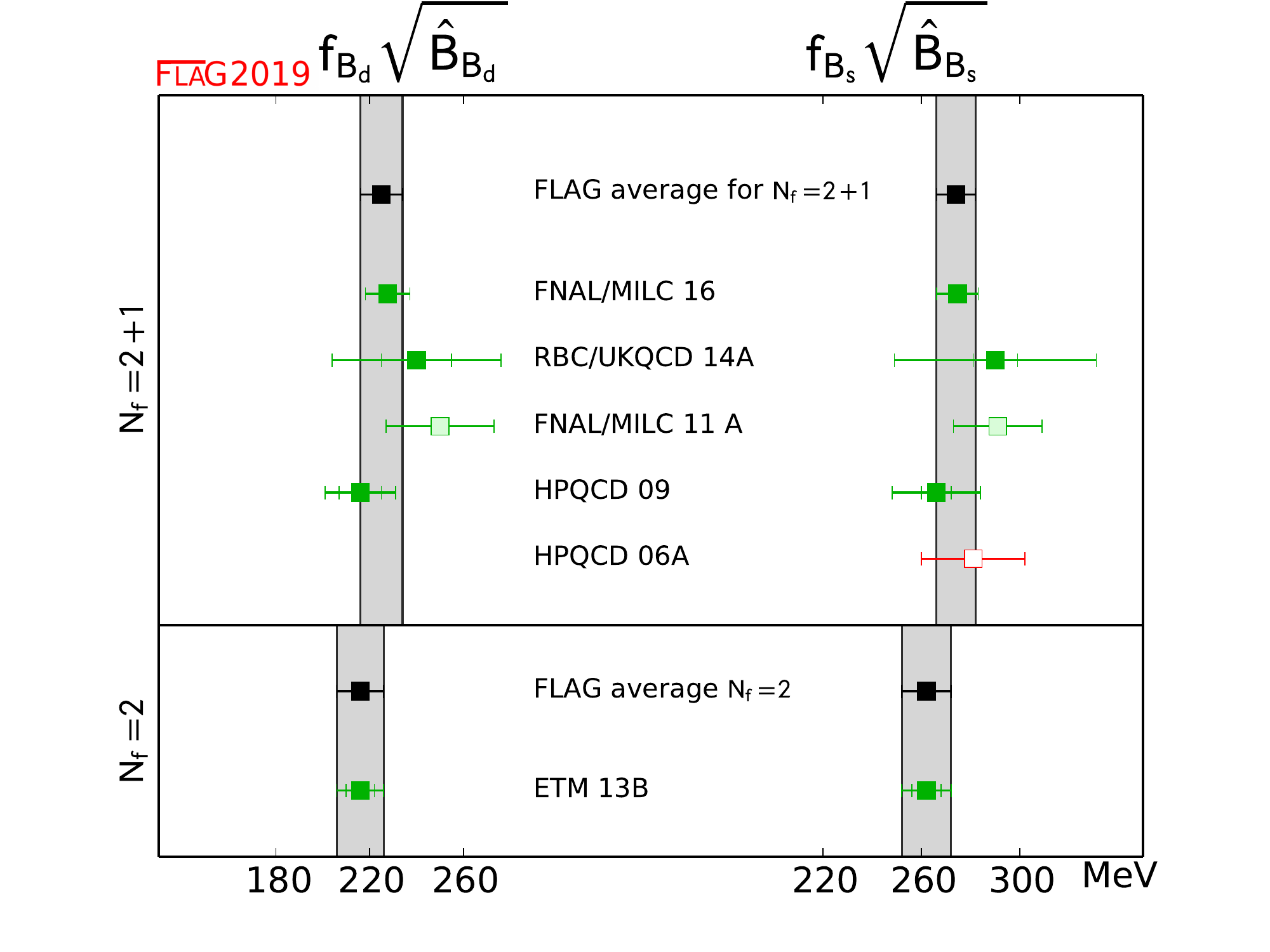}
\vspace{-5mm}
\caption{FLAG~\cite{Aoki:2019cca}
results for $f_{B_d}\sqrt{\hat B_{B_d}}$ and $f_{B_s}\sqrt{\hat B_{B_s}}$.
\label{fig:fBsqrtBB2}
}
\end{center}
\end{figure}

Briefly, it is conventional to define the effective Hamiltonian in terms of
eight operators, where only the first one appears in the standard model.
\begin{equation}
    \mathcal{H}_{\rm eff} = \sum_{i=1}^5 C_i \op_i^q + \sum_{i=1}^3 \tilde{C}_i \tilde{\op}_i^q,
    \label{eq:heff}
\end{equation}
The first operator is given by
\begin{equation}
\op_1^q =\bar{b}^\alpha\gamma_\mu L q^\alpha \, \bar{b}^\beta\gamma_\mu L q^\beta 
\end{equation}
and its matrix element is conventionally defined as~\cite{Gabbiani:1996hi}
\begin{equation}
 \me{1}(\mu) = c_1 f^2_{B_q} M^2_{B_q} B^{(1)}_{B_q}(\mu) ,
    \label{eq:Bq_1}
\end{equation}
where the last factor is called the bag parameter and would be 1 in the
vacuum saturation approximation.  It is also convenient to define a 
renormalization-group-invariant bag parameter 
$\hat B^{(1)}_{B_q}$:
\begin{equation}
    \hat B^{(1)}_{B_q} = \alpha_s (\mu)^{-\gamma_0/(2\beta_0)}
        \left[1 + \frac{\alpha_s (\mu)}{4\pi} \left( \frac{\beta_1\gamma_0 - \beta_0\gamma_1}{2\beta_0^2}
        \right) \right] \ B^{(1)}_{B_q}(\mu)
    \label{eq:RGI}
\end{equation}
In the standard model, we can express the difference in the masses of the
mixing eigenstates as
\begin{equation}
    \Delta M_q = \frac{G_F^2 m_W^2 M_{B_q}}{6\pi^2}\, S_0 (x_t )\, \eta_{2B} \, |V_{tq}^* V_{tb}|^2\,
        f^2_{B_q} \hat B^{(1)}_{B_q} .
    \label{eq:DM}
\end{equation}
Details can be found in Ref.~\cite{Bazavov:2016nty}.  The key point is
that the measurable mass difference is proportional to $|V_{tq}|^2$ and
the bag parameter (or that bag parameter multiplied by the square of
the decay constant).  Figure~\ref{fig:fBsqrtBB2} contains the 
FLAG~\cite{Aoki:2019cca}
summary of $f_{B_d}\sqrt{\hat B_{B_d}}$ and 
 $f_{B_s}\sqrt{\hat B_{B_s}}$.  There are three results that are
averaged for $N_f=2+1$ and one result for $N_f=2$.  
FLAG does not calculate the impact of these results on the determination
of the CKM matrix elements $V_{td}$ and $V_{ts}$; however, 
Ref.~\cite{Bazavov:2016nty} does.  That paper also contains results for 
BSM operators.

Using experimental results for $B_{(s)}$, mixing, it is found in 
Ref.~\cite{Bazavov:2016nty} that
\begin{eqnarray} 
        |V_{td}| & = & 8.00(33)(2)(3)(8) \times 10^{-3},  \label{eq:Vtd} \\
        |V_{ts}| & = & 39.0(1.2)(0.0)(0.2)(0.4) \times 10^{-3},  \label{eq:Vts} \\
        |V_{td}/V_{ts}| & = & 0.2052(31)(4)(0)(10), \label{eq:VtdOverVts}
\end{eqnarray} 
where in each case the first error is from the error in the lattice
mixing matrix element, the second error comes from the error in the experimental
mass difference, the third error comes from errors in parameters used
in  Eq.~(\ref{eq:DM}), and the final error is from the lack of charm quarks
in the sea.  One can see that the lattice errors are dominant, even in the 
ratio of CKM matrix elements in which the hadronic uncertainties are
suppressed.  The errors on each quantity are 4.3\%, 3.2\%, and 1.6\%, from
Eq.~\ref{eq:Vtd} to Eq.~\ref{eq:VtdOverVts}, respectively.

\section{Summary}

Table~\ref{tab:summary} shows the value of the magnitude of
each CKM matrix element except $V_{tb}$.  The second column has the value,
and the third, the percentage error.  We see that we have sub percent
accuracy in the first row, but the other elements have errors ranging
from 1.4\% to 4.3\%.  The final column indicates the source of each
result.  This is particularly pertinent for $|V_{ud}|$ and  $|V_{us}|$
as we have already remarked on the tensions with unitarity for the first
row, and what happens if one considers one leptonic and semileptonic
decays without input from nuclear $\beta$-decay on $|V_{ud}|$.

\begin{table}[th]
\begin{center}
\begin{tabular}{lccc}
\hline\hline
Quantity & value & percentage error & Comment \\
\hline
$|V_{ud}|$ & 0.9737(16) & 0.16 & FLAG (with unitarity)\\
$|V_{ud}|$ & 0.9669(34) & 0.35 & FNAL/MILC ($\rm K_{l2}$ \& $\rm K_{l3}$ )\\
$|V_{us}|$ & 0.2249(7) & 0.31 & FLAG (with unitarity)\\
$|V_{us}|$ & 0.22333(61) & 0.27 & FNAL/MILC ($\rm K_{l2}$ \& $\rm K_{l3}$)\\
$|V_{cd}|$ & 0.2219(43) & 1.9 & FLAG (2+1+1)\\
$|V_{cs}|$ & 1.002(14) & 1.4 & FLAG (2+1+1)\\
$|V_{ub}|\times 10^3$ & 3.76(14) & 3.7 & FLAG (BGL, combined)\\
$|V_{cb}|\times 10^3$ & 41.47(70) & 1.7 & FLAG (BGL, combined)\\
$|V_{td}|\times 10^3$ & 8.00(34) & 4.3 & FNAL/MILC\\
$|V_{ts}|\times 10^3$ & 39.0(1.3) & 3.2 & FNAL/MILC
\end{tabular}
\end{center}
\caption{Summary of CKM matrix elements determined with input from lattice
QCD.
\label{tab:summary}
}
\end{table}

\section{Prospects}

Yogi Berra is purported to have said, ``It is tough to make predictions, 
especially about the future.''  I trust you will keep that in mind as you read
this section.

We have seen that in a number of quantities the theory error is the limiting
factor in determining a CKM matrix element.  Even in cases for which the
experimental and theoretical errors are currently
comparable, we can expect that 
BESIII, Belle II, and LHCb will reduce the experimental errors.
The Belle II Physics Book~\cite{Kou:2018nap} details what is expected to be
accomplished at Belle II in many topics.  In order to improve
the future precision of CKM matrix elements, it is essential for increased
theoretical precision.  It is assumed in this work that there will be
a factor of five improvement in errors from lattice QCD in ten years.
Figure 85 of Ref.~\cite{Kou:2018nap} predicts the error on $V_{ub}$ from 
the study $B\to \pi \ell\nu$, taking into account both increased integrated
luminosity and presumed improvement in lattice QCD precision in five and 
ten years.  The figure makes clear how important it is to improve our
calculations to make the best use of data coming from Belle II.  Failure
to improve the theoretical input could increase the error on $V_{ub}$ by a
factor of two or more.

There has also been a recent report on future prospects of 
the LHC~\cite{Cerri:2018ypt}
where we can look forward to both the high-luminosity and higher 
energy enhancements of the machine.  This report assumes a factor of three
improvement in theoretical precision.  A white paper from USQCD on flavor
physics~\cite{Lehner:2019wvv} will also interest the reader.  We see that
there is both a need and an expectation that the lattice QCD community
will continue to improve our calculations.  

\section{Conclusions}
Over the past few years, there has been very significant progress in using
lattice QCD to calculate standard model parameters such
as quark masses, the strong coupling $\alpha_s$, and matrix elements need 
to determine the CKM matrix.  
A number of quantities are now available at the sub-percent
level, and we expect the precision to increase by a factor of 3--5 over
the next 5--10 years for form factors.  
Thus, the interplay between theory and experiment
will provide more and more stringent tests of the Standard Model (and,
perhaps, evidence of new physics).  We are getting to the level
of precision at which electromagnetic corrections are important.
The Rome-Southampton group has shown leadership in the area and
their work was presented in the parallel sessions.

Finally, BESIII, Belle II, and LHCb all have a large role to play in the
future of flavor physics.  I, for one, can hardly wait for their new results!

\section*{Acknowledgments}
I am grateful to the local organizers and the International Advisory Committee
for offering me the opportunity to give this talk.  I owe a great
debt to my friends and
collaborators in the Fermilab Lattice and MILC Collaborations.  
I am thankful to all the members of the Flavour Lattice Averaging Group,
whose work I have relied upon.  In particular, working with the other members
of the two working groups that deal with decay of hadrons containing charm
or bottom quarks, and $B$ meson mixing has been a pleasure.
Thanks to Zech Gelzer for preparing Fig.~\ref{fig:Vub}.
Finally, this work was supported by the U.S. Department of Energy
through grant DE-SC0010120.

\end{document}